\documentclass[12pt]{article}
\usepackage{amsfonts,amsmath,amssymb}
\usepackage{graphicx}

\newfont{\twelvemsb}{msbm10 scaled\magstep1}
\newfont{\eightmsb}{msbm8}
\newfam\msbfam
\textfont\msbfam=\twelvemsb \scriptfont\msbfam=\eightmsb
\catcode`\@=11
\def\Bbb{\ifmmode\let\next\Bbb@\else
\def\next{\errmessage{Use \string\Bbb\space only in math mode}}\fi\next}
\def\Bbb@#1{{\fam\msbfam{{#1}}}}

\newcommand{\be}{\begin{equation}}
\newcommand{\ee}{\end{equation}}
\newcommand{\ba}{\begin{eqnarray}}
\newcommand{\ea}{\end{eqnarray}}

\newcommand{\m}{\mathcal}

\newcommand{\nn}{\nonumber}
\newcommand{\q}{\theta}

\def\[{\left[}
\def\]{\right]}
\def\({\left(}
\def\){\right)}
\def\[{\left[}

\usepackage[usenames,dvipsnames]{color}

\begin{document}

\sloppy
\renewcommand{\thefootnote}{\fnsymbol{footnote}}
\newpage
\setcounter{page}{1} \vspace{0.7cm}
\vspace*{1cm}
\begin{center}
{\bf Fermions and scalars in $\mathcal{N} = 4$ Wilson loops at strong coupling and beyond} \\
\vspace{.6cm} {Alfredo Bonini$^{a}$, Davide Fioravanti $^a$, Simone Piscaglia $^{b}$, Marco Rossi $^c$
}
\footnote{E-mail: bonini@bo.infn.it, fioravanti@bo.infn.it, piscaglia@th.phys.titech.ac.jp, rossi@cs.infn.it} \\
\vspace{.3cm} $^a$ {\em Sezione INFN di Bologna, Dipartimento di Fisica e Astronomia,
Universit\`a di Bologna} \\
{\em Via Irnerio 46, 40126 Bologna, Italy}\\

\vspace{.3cm} $^b$ {\em Department of Physics,
Tokyo Institute of Technology,
Tokyo 152-8551, Japan}

\vspace{.3cm} $^c${\em Dipartimento di Fisica dell'Universit\`a
della Calabria and INFN, Gruppo collegato di Cosenza} \\
{\em Arcavacata di Rende, Cosenza, Italy} \\
\end{center}
\renewcommand{\thefootnote}{\arabic{footnote}}
\setcounter{footnote}{0}
\begin{abstract}
{\noindent We study the strong coupling behaviour of null polygonal Wilson loops/gluon amplitudes in $\mathcal{N} = 4$ SYM, by using the OPE series and its integrability features. For the hexagon we disentangle the $SU(4)$ matrix structure of the form factors for fermions, organising them in a pattern similar to the Young diagrams used previously for the scalar sector \cite{BFPR2,BFPR3}. Then, we complete and extend the discussion of \cite{BFPR1} by showing, at strong coupling, the appearance of a new effective particle in the series: the fermion-antifermion bound state, the so-called meson. We discuss its interactions in the OPE series with itself by forming (effective) bound states and with the gluons and bound states of them. These lead the OPE series to the known $AdS_5$ minimal area result for the Wls, described in terms of a set of TBA-like equations. This approach allows us to detect all the one-loop contributions and, once the meson has formed, applies to $\mathcal{N}=2$ Nekrasov partition function via the parallel meson/instanton (in particular, they share the mechanism by which their bound states emerge and form the TBA node). Finally, to complete the strong coupling analysis, we consider the scalar sector for any polygon, confirming the emergence of a leading contribution from the non-perturbative theory on the sphere $S^5$.
}

\end{abstract}
\vspace{6cm}

\newpage


\tableofcontents

\index{}
\newpage

\section{Introduction and summary}
\label{intro}
\setcounter{equation}{0}

In the realm of the supersymmetric gauge theories a special role is played by ${\cal N}=4$ Super Yang-Mills (SYM), with gauge group $SU(N_c)$ and dimensionless coupling constant $g_{YM}$.
The theory indeed appears at one side of the most known example of AdS/CFT correspondence \cite{MGKPW1,MGKPW2,MGKPW3}, i.e. the duality between type IIB superstring theory on $AdS_5 \times S^5$ and ${\cal N}=4$ SYM on the boundary of $AdS_5$, the $4d$ Minkowski spacetime.

Of particular interest is the planar limit $N_c\rightarrow \infty$, in which we also send $g_{YM}\to 0$ keeping the 't Hooft coupling $\lambda\equiv N_c g_{YM}^2$ fixed. The new coupling $\lambda$ is the only parameter of the planar theory and in literature it is often represented as $\lambda = 16 \pi ^2 g^2$. The planar $\mathcal{N}=4$ SYM shows remarkable connections with $1+1$ dimensional integrable models \cite{MZ, BS1,BS2,BS3,BS4,BS5,BES,TBA1,TBA2,TBA3,QSC1,QSC2}, which have allowed a better comprehension and partial proof of the aforementioned correspondence.

Historically, integrability was discovered when dealing with the spectral problem, \emph{i.e.} the computation of the anomalous dimension of local gauge invariant operators. More recently, it played an important role also in the evaluation of null polygonal Wilson loops (Wls). They have been proposed to be dual to gluon scattering amplitudes \cite{AM-amp, DKS, BHT}, which makes them even more interesting. Currently, this correspondence has been widely tested both at weak and strong coupling, so that it can be considered a well-established fact.

In any conformal quantum field theory, the Operator Product Expansion (OPE) technique can be applied, besides to the usual product of local operators, also to the null polygonal Wilson loops \cite{Anope}. This method has an intrinsic non-perturbative nature and recalls the Form Factor (FF) Infra-Red (IR) spectral series of the correlation functions in integrable quantum field theories, where the operator is a specific {\it conical twist} field \cite{Knizhnik:1987xp, CCAD, BSV1, Bel-Tw1, Bel-Tw2,  CDF}. The OPE series was developed for the Wls in $\mathcal{N}=4$ SYM by employing the underlying integrability of the theory, which manifests itself in the flux-tube, dual to the Gubser, Klebanov and Polyakov (GKP) \cite{GKP} string. The connection with the spin chain picture was investigated in \cite{Bel-Ope,Bel-Qua} and the OPE series for the Wilson loop was pushed forward in a series of papers \cite{BSV1,BSV2,BSV3}. Briefly, the proposal was to write the expectation values of Wls as an infinite sum over intermediate excitations on the GKP string vacuum. In specific, the excitations appearing in the series are multiparticle states: gluons and bound states, fermions, antifermions and, finally, scalars. Therefore, in order to pursue this strategy, one needs to know the dispersion laws of the GKP string excitations \cite{Basso} and the $2d$ scattering factors \cite{FRO6, Basso-Rej, FPR1} between them\footnote {On the string theory side, the worldsheet scattering matrices have been computed in \cite{BB1, BB2} and they are in perfect agreement with results of \cite {FPR}.}.

The validity of the OPE series has been checked, by explicit computations, in two different regimes. These are the weak ({\it e.g.} \cite{BSV2,P1,P2,DP}) and the strong coupling ({\it e.g.} \cite{BSV3,BSV4,BFPR2, BFPR1, Bel1509,ISS}) limits, where comparisons against gauge or string theory results, respectively, have been successfully performed. As concerns the strong coupling limit, string theory has so far given only the leading order as minimisation of the worldsheet area living in the $AdS_5$ space and insisting on the polygon in $4d$ Minkowski \cite{Anope,TBuA,YSA, Hatsuda:2010cc}. On the gauge side, this contribution is represented by gluons and fermions in the OPE series. On the other hand, the scalar excitations provide a non-perturbative correction eluding the minimal area argument \cite{BSV4}. This effect comes from the string dynamics on the sphere $S^5$ and it is described by an $O(6)$ correlation function: for the hexagon it has been extensively studied in \cite{Belsca,BFPR2, BFPR3}.

In \cite{BFPR1, FPR}, the re-summation of leading contributions by gluons and fermions/antifermions to the OPE series at strong coupling was performed and agreement with the string result was found. For the fermions, the procedure was based on the (unproven) hypothesis that they contribute, in this specific limit, not as single particles but through effective bound states fermion-antifermion ($f\bar{f}$), proposed and discussed in \cite{BSV3,FPR}, which we called mesons. The latter are singlets under $SU(4)$ and the multiparticle form factors enjoy a simple factorisable form, which is a key property allowing the re-summation. In \cite{BFPR1} we have been able to prove that the meson hypothesis is indeed correct up to two couples $f\bar{f}$ ($n=2$). What has been missing so far is an argument for any $n$, the main obstacle being the complicated $SU(4)$ matrix structure when $n$ couples $f\bar{f}$ are involved. In this paper we fill the gap, definitely showing that the strong coupling limit of $n$ couples $f\bar{f}$ is described by $n$ mesons, whose matrix part is trivial, namely singlets under $SU(4)$. In turn, mesons form bound states by themselves and, upon resummation, reconstruct the central node of the TBA-like equations for the Wls/amplitudes. In addition, in \cite{BFPR1} a parallel between the mesons in $W$ and the instanton in $\mathcal{N}=2$, encoded in the Nekrasov function $\mathcal{Z}$ \cite{Nek}, has been revealed in the analysis of the $n=2$ $f\bar{f}$ contribution. Here we address the issue in more generality, with a particular focus on the so-called Nekrasov-Shatashvili (NS) limit for $\mathcal{N}=2$, associated to the strong coupling regime of $W$. This analogy also sheds some additional light on the one-loop contributions coming from the fermion sector: the part due to the mesons should follow the same pattern of the subleading correction to the NS limit, computed in \cite{BouFio1,BouFio2}. 

Concerning scalars, their contribution is purely non-perturbative from the string point of view and it is due to the dynamics on the five-sphere $S^5$. The OPE series for them is nothing but a form factors series of a $(N-4)$-point function in the $O(6)$ non-linear sigma-model. In the strong coupling limit the mass is exponentially suppressed $m\sim e^{-\frac{\sqrt{\lambda}}{4}}$, so that the strong coupling corresponds to the short-distance regime for the correlator. The expected power-law in the UV limit gives a contribution to the Wls of the same order as the classical one \cite{BSV4}. In \cite{BFPR2, BFPR3} the conformal limit of the hexagonal OPE series has been analysed by expanding the logarithm in terms of the connected functions, confirming the earlier proposal. In this paper we extend the argument to the general polygon, enabling us to prove the non-perturbative enhancement. In addition, the expansion over the multi-connected functions allows us to find a remarkable recursion formula between different polygons which, under some simple assumptions, reproduce the scaling proposed in \cite{BSV4} for the strong coupling limit.

The plan of the paper is the following. In Section \ref{fermi}, we start the analysis of the fermion contribution to the Wilson loop. We first deal with the matrix structure $\Pi_{mat}^{(n)}$ of the fermionic transitions (form factors), which does not depend on the coupling. It can be represented as a multiple integral over the isotopic roots associated to the residual $SU(4)$ R-symmetry of the underlying spin chain. Along the same line of the Young tableaux method developed in \cite{BFPR2,BFPR3}, we evaluate systematically the multiple residues and get an answer in terms of rational functions, formula (\ref{Pimatfinal}). The polar structure (\ref{Mat-fer}) of the matrix factor is also determined, the remaining non trivial information being encoded in certain polynomials $P^{(n)}$, which enjoy some nice relations listed in Appendix \ref{appA}.

In Section \ref{ffcont}, we work out the contribution of $n$ couples $f\bar{f}$ by means of the properties of the matrix part previously obtained. We compute the integrals over the antifermionic rapidities $v_i$ in the strong coupling limit by multiple residues, which results in (\ref{SingMes}). This is a series over effective particles (mesons) composed by one fermion and one antifermion, whose measure and pentagon transition are defined in terms of the fermionic ones and are valid at any coupling. However, in the strong coupling limit it is the only leading contribution to the original OPE series involving fermions. In order to prove this fact, a fundamental role is played by the properties of the polynomials $P^{(n)}$, listed in Appendix \ref{appA}.

In Section \ref{Mes-Ins} we throw a parallel between the series over mesons $W_M$ and the Nekrasov instanton partition function $\mathcal{Z}$, which encodes the effects of instantons in certain $\mathcal{N}=2$ gauge theories. It depends on two deformation parameters $\epsilon_1$, $\epsilon_2$, but for our purpose we consider only $\epsilon_2\equiv \epsilon$. The strong coupling limit in $\mathcal{N}=4$ corresponds to the $\epsilon\to 0$ NS limit for the partition function $\mathcal{Z}$, where a solution in terms of a TBA-like equation and the corresponding Yang-Yang functional is known \cite{NekSha}\footnote{For a detailed derivation, see also \cite{Bou,MenYang}.}. The interaction term in $\mathcal{Z}$ is composed by a long-range and a short-range part. They can be efficiently dealt with, respectively, by a Hubbard-Stratonovich transformation and the Fredholm determinant technique, whose combined application allows us to obtain a nice representation of $\mathcal{Z}$ (\ref{ZFre}). This is done in a detailed manner in Appendix \ref{NekApp}. From this representation the NS limit is straightforward and the TBA-like equation (\ref{TBAlikeZ}) is easily reproduced. In the main text, the very same procedure is applied to the meson series $W_M$, which is recast in the form (\ref{WFre}). Employing the strong coupling limit, we find (\ref{WLi2}) that agrees with the result in \cite{FPR} where, performing the saddle point approximation, the (central node of) TBA-like equations for the amplitudes (\ref{TBAamp}) and the associated Yang-Yang functional (\ref{YYamp}) has been obtained. In \cite{FPR}, to reproduce the TBA-like equation, the mesons and their bound states has been put by hand in the OPE series, see (\ref{BoundMes}). In this paper we obtained the same results directly from the OPE containing fermions and antifermions, definitively proving the validity of the meson hypothesis. As a completion of our analysis, we include the gluons in the treatment to obtain the full set of TBA-like equations.
Subsection \ref{oneloop} provides a sketch of the subleading corrections in the fermionic sector, organised by their different origins.

In Section \ref{scalars}, we extend the argument in \cite{BFPR2, BFPR3} to the general polygon $N>6$ by expanding the logarithm in multi-connected functions, mimicking what has been done for the hexagon previously. This makes possible to extract a factor $\sqrt{\lambda}$ in front of any term of the series and prove the emergence of the non-perturbative enhancement pushed forward in \cite{BSV4}. Furthermore, the peculiar properties of the multi-connected functions imply an interesting recursion relation among the polygons with different number of sides. This enables us to investigate the strong coupling/small distance limit of the Wilson loop and successfully compare with the result by \cite{BSV4}. Many technical steps are arranged in the dedicated Appendix \ref {scal}.

\section{The fermion contribution to the hexagonal Wilson loop}
\label{fermi}
\setcounter{equation}{0}

\begin{figure} [htbp]
\centering
\includegraphics[width=0.85\textwidth]{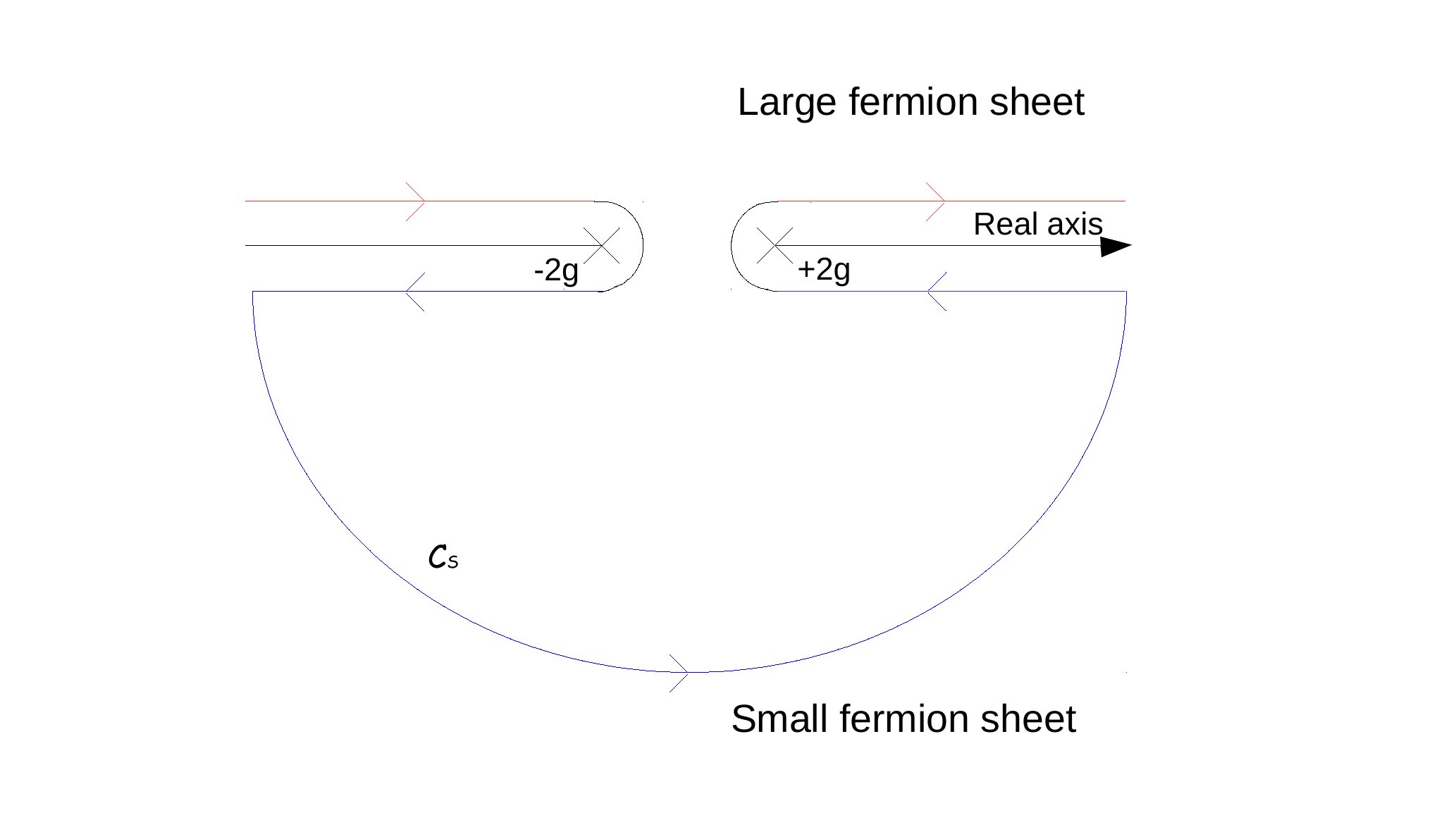}
\caption{The integrations on fermionic rapidities are performed along the path depicted in the figure above, going through the branch cut $[-2g,+2g]=[-\frac{\sqrt{\lambda}}{2\pi},\frac{\sqrt{\lambda}}{2\pi}]$; at strong coupling, the contribution coming from large fermions (corresponding to the red part of the contour, $\m{C}_{L}$) is negligible with respect to the small fermion contribution. Therefore, throughout this section the integrations will be restricted to the path $\m{C}_{S}$, corresponding to the blue curve, completely lying in the small fermion sheet.}
\label{Csmall}
\end{figure}

As a first step we review the contribution that fermion excitations bring to the null hexagonal bosonic Wilson Loop -- dual to the maximum helicity violating (MHV) six gluon scattering amplitude -- so to highlight its key features and set our notations throughout the paper. We make use of the OPE series \cite{BSV1,BSV2}, relying on pentagon amplitudes $P$  (which are form factors of a specific conical twist operator $\mathcal{P}$) as building blocks to represent the Wl as a sum over flux-tube excitations.
Since we focus on MHV amplitudes, only intermediate states which are singlets under the SU(4)(residual) R-symmetry, are taken into account in the sum. As discussed in Appendix \ref {appB}, the singlet condition, restricted to fermions, reads (\ref {ferm-singl}) $N_f=N_{\bar{f}}$ mod $4$. Nevertheless we focus exclusively on the configuration $N_f=N_{\bar{f}}=n$, since it is the only one we need to reconstruct the classical string results, as shown in \cite{BSV3,FPR, BFPR1}. The fermion contribution to the hexagonal Wl is conveniently decomposed as
\be\label{WL}
W_f=\sum_{n=0}^{\infty} W^{(n)}_f \, ,
\ee
in terms of $W^{(n)}_f $, which represents the contribution due to $n$ couples of fermions-antifermions,
\ba\label{WLferm}
&& W^{(n)}_f =\frac{1}{n! n!}\int _{\m{C}}\prod_{k=1}^n \left[\frac{du_k}{2\pi}\frac{dv_k}{2\pi}
\,\mu_f(u_k)\mu_f(v_k)\,e^{-\tau E_f(u_k)+i\sigma p_f(u_k)}\times \right.\\
&& \left. \times e^{-\tau E_f(v_k)+i\sigma p_f(v_k)}\right]
\Pi_{dyn}^{(n)}(\{u_i\},\{v_j\})\,\Pi_{mat}^{(n)}(\{u_i\},\{v_j\})\ , \nn
\ea
with $\{u_k\}$ ($\{v_k\}$) the sets of fermions (antifermions) rapidities. The energy and momentum of a fermion or antifermion are $E_f(u)$ and $p_f(u)$, respectively, while the cross ratios $\tau$ and $\sigma$ fix the conformal geometry of the hexagon. A third cross ratio $\phi$, coupled to the helicity of the particles, plays no role here; it will enter the stage in Section \ref{glufer}, where gluons too will be taken into account. The multiparticle transitions,
\be\label{split}
\Pi_{dyn}^{(n)}(\{u_i\},\{v_j\})\,\Pi_{mat}^{(n)}(\{u_i\},\{v_j\}) \ ,
\ee
are factorised into the product of a dynamical and a (coupling independent) matrix part
and represent the squared form factors of the operator $\mathcal{P}$, summed over the $SU(4)$ symmetry indices of the fermions\footnote {The matrix structures of the form factors of $\mathcal{P}$ have been clarified in \cite {Belmat}.}. That said, the dynamical part itself is factorized in terms of objects involving just two particles at once
\be\label{dyn}
\Pi_{dyn}^{(n)}(\{u_i\},\{v_j\}) = \displaystyle\prod_{i<j}^n\frac{1}{P^{(ff)}(u_i|u_j)P^{(ff)}(u_j|u_i)}
\frac{1}{P^{(ff)}(v_i|v_j)P^{(ff)}(v_j|v_i)}\displaystyle\prod_{i,j=1}^n \frac{1}{P^{(f\bar f)}(u_i|v_j)P^{(f\bar f)}(v_j|u_i)} \, ,
\ee
where $P^{(ff)}$ encodes the transition between particles of the same type ($i.e.$ fermion-fermion or antifermion-antifermion)
and $P^{(f\bar f)}$ the transition between a fermion and an antifermion. The function $P^{(ff)}(u|v)$ has a pole when $v=u$ and the relative residue,
\be\label{misura_fermione}
\mbox{Res}\,_{v=u}\,P^{(ff)}(u|v)=\frac{i}{\mu_f(u)} \ ,
\ee
determines the measure $\mu_f(u)$ \cite{BSV1} appearing in (\ref {WLferm}).
A crucial remark concerns the integrations in (\ref{WLferm}). The integration contour $\m{C}$ lies on a Riemann surface composed of two sheets, namely the large and small fermion sheet \cite{Basso}, connected by the branch cut $[-\frac{\sqrt{\lambda}}{2\pi},\frac{\sqrt{\lambda}}{2\pi}]$. Hence $\m{C}$ naturally splits into a section $\m{C}_{L}$, contained in the large fermion sheet, and $\m{C}_{S}$, lying on the small fermion sheet, as depicted in Figure\,\ref{Csmall}.
In the strong coupling $\lambda\rightarrow\infty$ limit (the regime we are mainly interested in), only the latter section contributes to the integral: in fact, in the large fermion sheet the particle energy $E_f$ scales as $\sim\sqrt{\lambda}$ (giant magnon regime), thus producing an exponentially dumped factor in (\ref{WLferm}). We refer to \cite{BSV2,Basso,FPR} for exhaustive explanations.

\subsection{The matrix factor}

The factor $\Pi^{(n)}_{mat}$, encoding the $SU(4)$ matrix structure, has an integral representation \cite{BSV4,BSV6} in terms of the auxiliary variables $a_i,\,b_i,\,c_i$, corresponding to the nodes of the $SU(4)$ Dynkin diagram\footnote{The connection between the number of particles in a singlet state and the Bethe equations for the $SU(4)$ spin chain is discussed in Appendix \ref{appB}.}. In a system composed of $n$ fermions with rapidities $u_i$ and $n$ antifermions with rapidities $v_i$, arranged in a configuration not charged under $SU(4)$, we have $n$ isotopic roots of each type and the matrix factor reads \cite{BSV6}
\ba\label{Pi_mat^ff}
\Pi_{mat}^{(n)}(\{u_i\},\{v_j\})=\frac{1}{(n!)^3}\int
\prod_{k=1}^n\left(\frac{da_k db_k dc_k}{(2\pi)^3}\right)
\,\frac{\displaystyle\prod_{i<j}^n g(a_i-a_j) g(b_i-b_j) g(c_i-c_j)}
{\displaystyle\prod_{i,j}^n f(a_i-b_j) f(c_i-b_j) \prod_{i,j}^n f(u_i-a_j) f(v_i-c_j)} \nn \, , \\
\,
\ea
where the integrations are performed on the whole real axis and
\be
f(u)=u^2+\frac{1}{4}, \quad g(u)=u^2(u^2+1)=f(u-\frac{i}{2})f(u+\frac{i}{2}) \ . \nn
\ee
A fruitful simplification occurs upon realising that the variables $a_i$ and $c_i$ appear symmetrically in (\ref{Pi_mat^ff}) and do not couple to each other. Therefore they can be integrated out separately, to give the equal factors
\be\label{int_a}
\mathcal{D}_{2n}(x_1,\ldots,x_{2n})\equiv
\int
\prod_{k=1}^{n}\frac{da_k}{2\pi}
\frac{\displaystyle\prod_{i<j}^{n} g(a_i-a_j)}
{\displaystyle \prod_{j=1}^{2n} \prod_{i=1}^{n} f(a_i-x_j)}=2n \frac{\delta _{2n}(x_1,\ldots,x_{2n})}{\prod \limits_{\stackrel {i,j=1}{i<j}} ^{2n} [ (x_i-x_j)^2+1]} \ .
\ee
Hence the matrix factor can be written as
\ba
\Pi_{mat}^{(n)}(\{u\},\{v\}) = \frac{1}{(n!)^3}\int\prod_{k=1}^n\frac{db_k}{2\pi}\prod_{i<j}^n g(b_{ij})
\m{D}_{2n}(b_1,\dots,b_n,u_1,\dots,u_n)\m{D}_{2n}(b_1,\dots,b_n,v_1,\dots,v_n)  \,.\nn
\ea
The function $\delta_{2n}$ in (\ref{int_a}) is symmetric under permutations of its arguments $x_i$, depends on them only through their differences $x_i-x_j=x_{ij}$ and enjoys an appealing expression in terms of the Pfaffian of the antisymmetric $2n \times 2n$ matrix $\mathfrak{D}$ \cite{BFPR3}:
\be\label{deltaPf}
\delta_{2n}(x_1,\ldots ,x_{2n})=
\frac{2^n n!}{2n}\displaystyle\prod_{i<j}\frac{x_{ij}^2+1}{x_{ij}}\textrm{Pf} \, \mathfrak{D} \ ,
\quad \mathfrak{D}_{ij}=\left(\frac{x_{ij}}{x_{ij}^2+1}\right) \ .
\ee
One can easily realise that $\delta_{2n}$ is a polynomial: indeed, the symmetry under exchange of its arguments rules out the apparent single poles for coinciding variables in (\ref {deltaPf}).
For further properties of $\delta_{2n}$ we address the reader to Appendix \ref{Appdelta}.
Eventually, (\ref{Pi_mat^ff}) turns to the form
\ba\label{ff-int}
\Pi_{mat}^{(n)}(\{u_i\},\{v_j\}) &=&
\frac{4n^2}{(n!)^3}\frac{1}{\displaystyle\prod_{i<j}^n(u_{ij}^2+1)(v_{ij}^2+1)}
\int\prod_{k=1}^n\frac{db_k}{2\pi} \prod_{i<j}^n\left(\frac{b_{ij}^2}{b_{ij}^2+1}\right)\cdot \nn\\
& \cdot &
\frac{\delta_{2n}(b_1,\dots,b_n,u_1,\dots,u_n)\delta_{2n}(b_1,\dots,b_n,v_1,\dots,v_n)}{\displaystyle\prod_{i,j=1}^n[(b_i-u_j)^2+1][(b_i-v_j)^2+1]}
\ea
which paves the way for a systematic evaluation by residues: indeed, the matrix factor $\Pi_{mat}^{(n)}$ can be computed as a sum over specific residue configurations, following the strategy already carried out for scalars in \cite{BFPR2,BFPR3}.
Some preliminary remarks about the pole structure of (\ref{ff-int}) are in order:
\begin{enumerate}
	\item the double zeroes for $b_i=b_j$ prevent singularities for coinciding $b$'s: for example, if we take the residue $b_1=u_i+i$ in the first integral, we do not need to consider the residues in $b_{j\neq 1}=u_i+i$ any more, as they are vanishing;
	\item poles due to factors $\frac{1}{b_{ij}^2+1}$ play no role;
	\item one needs to evaluate the residues for poles of the type $b_k-u_j=i$ or $b_k-v_j=i$ only, and at most once for a given physical rapidity:
	  as an example, if we compute a residue for $b_1=u_j+i$, poles at $b_{k\neq 1}=u_j+i$ or $b_{k\neq 1}=u_j+2i$ do not occur.
\end{enumerate}
The remarks $(2)$ and $(3)$ can be understood by noting that in (\ref{ff-int}) half of the entries of the $\delta_{2n}$ polynomials correspond to the $n$ integration variables $b_j$, whereas the remaining $n$ are fermionic rapidities, $i.e.\ u_k$ or $v_k$, and in addition to that, the property (\ref{delta_align}) justifies our claims.

In a diagrammatic language, the considerations above mean that one needs to consider Young diagrams\footnote{Although, strictly speaking, the use of the term is improper.} where $n$ boxes, each one corresponding to the contribution of a single pole, shall be arranged into an array with $2n$ entries, $i.e.$ related to the $n$ fermionic plus the $n$ antifermionic rapidities\footnote{This represents a difference with respect to the scalar case. Indeed, for fermion excitations the number of rapidities is twice the number of integrations. For scalars instead, the number of rapidities matches the number of integrations.},
paying attention not to place two boxes on the same position (differently said, two poles with the same real coordinate cannot coexist).
Under these prescriptions, the matrix factor (\ref{ff-int}) can be written as
\be\label{ff-Y}
\Pi_{mat}^{(n)}(\{u_i\},\{v_j\})=\frac{4n^2}{(n!)^2 2^n}\frac{1}{\displaystyle\prod_{i<j}(u_{ij}^2+1)(v_{ij}^2+1)}Y_n(\{u_i\},\{v_j\})
\ee
where $Y_n$ denotes a sum over diagrams
\be\label{sumY}
Y_n(\{u_i\},\{v_j\})=\sum_{l_1+\ldots +l_{2n}=n, l_i=0,1} (l_1,\ldots,l_{2n}) \,;
\ee
and, in each diagram $(l_1,\ldots,l_{2n})$, the first $n$ entries are related to the fermion rapidities $u_i$, while the remaining $n$ correspond to antifermions $v_i$. More explicitly, $l_k=1$ (for $k\leq n$) means that the residue for the pole $b_j-u_k=i$ has been evaluated in (\ref{ff-int}), for some $b_j$, whereas $l_k=1$ (with $k> n$) is associated to $b_j-v_{k-n}=i$.\\
The number of diagrams appearing in the sum (\ref{sumY}) amounts to $(2n)!/(n!)^2$, corresponding to the non-equivalent permutations of the $l_k$'s entries of $(l_1,\ldots,l_{2n})$. Actually, all the diagrams can be obtained from a single one, say for simplicity
\ba\label{Perm}
(1,\ldots ,1,0,\ldots ,0)= \frac{\delta_{2n}(u_1+i,\ldots ,u_n+i,u_1,\ldots ,u_n)\delta_{2n}(u_1+i,\ldots ,u_n+i,v_1,\ldots ,v_n)}{\displaystyle\prod_{i<j}^n(u_{ij}^2+1)(u_{ij}^2+4)\displaystyle\prod_{i,j=1}^n (u_i-v_j)(u_i-v_j+2i)} \,,
\ea
by considering a suitable permutation of the $2n$ variables $u_i$ and $v_i$.
To get a different diagram, we need to change the positions of some \textbf{1}-entries in the array and permute the rapidities accordingly: we just need to pay attention that whenever a \textbf{1} is moved from the position $i\leq n$ to $n+j$ ($1\leq j\leq n$), we shall swap the two rapidities $u_i$ and $v_j$ in (\ref{Perm}), with the \textit{caveat} that the second half of arguments in the $\delta_{2n}$-polynomials must be held fixed (as they are already fixed inside the integrals in (\ref{ff-int})\,).
We denote as
\be\label{diag_gen}
(\boldsymbol{1}_k,\boldsymbol{0};\boldsymbol{0},\boldsymbol{1}_{n-k})\equiv (1,\ldots , 1_k,0,\ldots ,0_n,0,\ldots ,0_k,1,\ldots ,1)
\ee
the most general contribution (out of a permutation of the variables which does not mix the $u$ and the $v$ rapidities), with $k$ \textbf{1}'s placed on the left, corresponding to the variables $u_i$, and $n-k$ \textbf{1}'s on the right, representing the $v_i$.
Thanks to (\ref{delta201.1}), the diagram (\ref{diag_gen}) is given an explicit form
\ba\label{Genk}
(\boldsymbol{1}_k,\boldsymbol{0};\boldsymbol{0},\boldsymbol{1}_{n-k}) &=& \frac{\delta_{2k}(u_{1}+i,\ldots ,u_{k}+i,v_1,\ldots ,v_k)\delta_{2n-2k}(u_{k+1},\ldots ,u_n,v_{k+1}+i,\ldots ,v_n+i)}{\displaystyle\prod_{i,j=1}^k(u_i-v_j)(u_i-v_j+2i)\displaystyle\prod_{i,j=k+1}^n(u_i-v_j)(u_i-v_j-2i)}\cdot \nn\\
&\cdot & \displaystyle\prod_{i=1}^k\displaystyle\prod_{j=k+1}^n\frac{(u_{ij}-i)(v_{ij}+i)}{u_{ij}v_{ij}} \cdot 2^n\frac{[(n-1)!]^2}{(k-1)!(n-k-1)!} \ ,
\ea
which holds for any $k=1,\ldots,n-1$.
Finally, the matrix factor (\ref{ff-Y}) comes out as a sum of rational functions, where, for fixed $k$, we need to add together
$\binom{n}{k}^2$ contributions: each one of them follows from applying to (\ref{Genk}) a permutation $P$ on the fermionic variables $u_i$ and a permutation $Q$ on the antifermionic ones $v_i$, up to a normalisation factor to avoid over-counting.
The matrix factor $\Pi^{(n)}_{mat}$ eventually reads\footnote{In order for the $k=0$ and $k=n$ terms to make sense, one must substitute the factorial $(-1)!$ with $2$.}
\ba\label{Pimatfinal}
&&\Pi^{(n)}_{mat}(u_1, \ldots ,u_n,v_1,\ldots ,v_n) = \frac{4}{\displaystyle\prod_{i<j}^n(u_{ij}^2+1)(v_{ij}^2+1)}\sum_{k=0}^n\frac{1}{[(n-k)!(k)!]^2(k-1)!(n-k-1)!}\cdot \nn\\
&\cdot &\sum_P \sum_{Q}\frac{\delta_{2k}(u_{P_1}+i,\ldots ,u_{P_k}+i,v_{Q_1},\ldots ,v_{Q_k})\delta_{2n-2k}(u_{P_{k+1}},\ldots ,u_{P_n},v_{Q_{k+1}}+i,\ldots ,v_{Q_n}+i)}{\displaystyle\prod_{i,j=1}^k(u_{P_i}-v_{Q_j})(u_{P_i}-v_{Q_j}+2i)\displaystyle\prod_{i,j=k+1}^n(u_{P_i}-v_{Q_j})(u_{P_i}-v_{Q_j}-2i)} \cdot \nn\\
&\cdot & \displaystyle\prod_{i=1}^k\displaystyle\prod_{j=k+1}^n\frac{(u_{P_i}- u_{P_j}-i)(v_{Q_i}-v_{Q_j}+i)}{(u_{P_i}- u_{P_j})(v_{Q_i}-v_{Q_j})} \ .
\ea
For the sake of clarity, the simplest examples, $i.e.\ n=1$ and $n=2$, are portrayed below.\\
\medskip\\
\textbf{$\bullet $ One couple fermion-antifermion ($n=1$):}\\
In the simplest case $n=1$ only two diagrams contribute, $(1,0)$ and $(0,1)$, the former from computing the residue for the pole $b-u_1=i$,
$$
(1,0)=\frac{1}{(u-v)(u-v+2i)} \,,
$$
the latter for $b-v_1=i$,
$$
(0,1)=\frac{1}{(v-u)(v-u+2i)} \,.
$$
The expression (\ref{ff-Y}) amounts to
$
\Pi_{mat}^{(1)}(u,v)=2Y_1(u,v)=2\left[(1,0)+(0,1)\right]
$,
so we obtain:
\be
\Pi_{mat}^{(1)}(u,v)=\frac{4}{(u-v)^2+4} \,.
\ee
\medskip\\
\textbf{$\bullet $ Two couples fermion-antifermion ($n=2$):}\\
The $n=2$ case may be more clarifying. When computing the matrix factor
\be\label{Pi-Y}
\Pi_{mat}^{(2)}(u_1,u_2,v_1,v_2)=\frac{1}{(u_{12}^2+1)(v_{12}^2+1)}Y_2(u_1,u_2,v_1,v_2)
\ee
we take into account six distinct contributions, namely
\ba
Y_2(u_1,u_2,v_1,v_2)=(1,1,0,0)+(1,0,1,0)+(1,0,0,1)+(0,1,1,0)+(0,1,0,1)+(0,0,1,1) \ .\nn
\ea
The first term results from (\ref{Perm}), upon using (\ref{delta20}):
\ba\label{1100}
(1,1,0,0) &=&
\frac{\delta_{4}(u_1+i,u_2+i,u_1,u_2)\delta_{4}(u_1+i,u_2+i,v_1,v_2)}{\displaystyle(u_{12}^2+1)(u_{12}^2+4)\displaystyle\prod_{i,j}^2 (u_i-v_j)(u_i-v_j+2i)}= \\
&=& \frac{2\delta_4(u_1+i,u_2+i,v_1,v_2)}{\displaystyle\prod_{i,j=1}^2(u_i-v_j)(u_i-v_j+2i)} \nn
\ea
whereas $(0,0,1,1)$ is obtained from (\ref{1100}) by performing the substitution $(u_1,u_2)\leftrightarrow (v_1,v_2)$.
The diagram $(1,0,1,0)$ can be retrieved from the first line of (\ref{1100}) by exchanging $u_1\leftrightarrow v_1$:
\ba
(1,0,1,0) &=&
\frac{\delta_{4}(u_1+i,v_1+i,u_1,u_2)\delta_{4}(u_1+i,v_1+i,v_1,v_2)}{\displaystyle[(u_{1}-v_1)^2+1][(u_{1}-v_1)^2+4]}\cdot \\
&\cdot &\frac{1}{u_{12}(u_{12}+2i)v_{12}(v_{12}+2i)(u_1-v_2)(u_1-v_2+2i)(v_1-u_2)(v_1-u_2+2i)}= \nn\\
&=&\frac{4(u_1-u_2-i)(v_1-v_2-i)}{(u_1-v_2)(v_1-u_2)(u_1-u_2)(v_1-v_2)(u_1-v_2+2i)(v_1-u_2+2i)} \,.
\ea
Then, the remaining terms follow after a suitable permutation of the variables: $(1,0,0,1)$ results from exchanging $v_1\leftrightarrow v_2$, $(0,1,1,0)$ from $u_1\leftrightarrow u_2$ and $(0,1,0,1)$ after the exchange of both $(u_1,v_1)\leftrightarrow (u_2,v_2)$.
Summing up all the contributions, the matrix factor (\ref{Pi-Y}) amounts to
\be
\Pi_{mat}^{(2)}(u_1,u_2,v_1,v_2)=\frac{1}{((u_{1}-u_2)^2+1)((v_{1}-v_2)^2+1)}\frac{P^{(2)}(u_1,u_2,v_1,v_2)}{\displaystyle\prod_{i,j=1}^2((u_i-v_j)^2+4)}
\ee
(where the polynomial $P^{(2)}$ is displayed in (\ref{P1P2})\,), thus confirming the findings by \cite{BFPR1,Bel1509}.

\subsection{Residues of matrix factor and recursion formula}

In \cite{Bel1509} a recursion relation for matrix factors was found in the case when the number of fermions and antifermions differ by one.
Here we put forward an analogous formula, valid for equal number $n$ of fermions and antifermions, which expresses the residue of the matrix factor $\Pi_{mat}^{(n)}$ evaluated in $v_j=u_i+2i$, in relation to $\Pi_{mat}^{(n-1)}$, involving one less couple fermion-antifermion:
\be\label{ResPimat}
i \textit{Res}_{v_1=u_1+2i} \Pi_{mat}^{(n)}(u_1,\ldots ,u_n,v_1,\ldots ,v_n)= \frac{\Pi_{mat}^{(n-1)}(u_2,\ldots ,u_n,v_2,\ldots ,v_n)}{\displaystyle\prod_{j=2}^n(u_{1j}+i)u_{1j}(u_1-v_j+2i)(u_1-v_j+i)} \,.
\ee
This formula can be shown directly from the sum over configurations (\ref{ff-Y}), by considering the contribution of a single diagram (\ref{Genk}). The proof relies on the consideration that the pole $v_1=u_1+2i$ appears only in diagrams of the type $(1,\lbrace l_{i=2}^n \rbrace , 0, \lbrace l_{n+1}^{2n} \rbrace)$, which factorise into three distinct factors
\be
(1,\lbrace l_{i=2}^n \rbrace , 0, \lbrace l_{n+1}^{2n} \rbrace)=(1,0)\cdot (\lbrace l_{i=2}^n \rbrace , \lbrace l_{n+1}^{2n} \rbrace)\cdot M_{\lbrace l_i \rbrace} \ :
\ee
the two-particle diagram $(1,0)$ contains the pole; $(\lbrace l_{i=2}^n \rbrace , \lbrace l_{n+1}^{2n} \rbrace)$ represents one of the diagrams contributing to $\Pi_{mat}^{(n-1)}$; the mixing term $M_{\lbrace l_i \rbrace}$, whose expression can be derived from (\ref{Genk}), depends on both sets of variables and on the specific diagram ${\lbrace l_i \rbrace}$. When evaluating the residue around $v_1=u_1+2i$, the dependence on the diagram drops out and, upon summing over the set ${\lbrace l_i \rbrace}$, the matrix factor with a decreased number of particles $\Pi_{mat}^{(n-1)}$ shows up, resulting in (\ref{ResPimat}).\\
This kind of relation does not come unexpected from a physical ground, for it clearly alludes to the customary
form factor\footnote{Although, strictly speaking, here we are dealing with some square modulus of a form factor.}
axiom on kinematic poles.

\subsection{Polar structure and fermion polynomials}

For a system consisting of $n$ couples fermion--antifermion, the matrix factor enjoys the structure:
\be\label{Mat-fer}
\Pi_{mat}^{(n)}(\{u_i\},\{v_j\})=\frac{P^{(n)}(u_1,\dots,u_n,v_1,\dots,v_n)}
{\displaystyle\prod_{i<j}^n[(u_i-u_j)^2+1]\prod_{i<j}^n[(v_i-v_j)^2+1]\prod_{i,j=1}^n[(u_i-v_j)^2+4]} \,,
\ee
where $P^{(n)}(u_1,\dots,u_n,v_1,\dots,v_n)$ is a degree $2n(n-1)$ polynomial in fermion and antifermion rapidities $u_i,\,v_j$\,.
We write down the explicit expression for $P^{(n)}$ only in the simplest cases ($n=1,2$), since it rapidly grows cumbersome as $n$ increases:
\ba\label{P1P2}
P^{(1)}(u_1,v_1) &=& 4 \\
P^{(2)}(u_1,u_2,v_1,v_2) &=& 4[24 + 3((u_2-v_1)^2+6)((u_1-v_2)^2+6) + \nn\\
&+&   3((u_1-v_1)^2+6)((u_2-v_2)^2+6) +((u_1-u_2)^2+4)((v_1-v_2)^2+4)] \ . \nn
\ea
The proof of the general polar structure (\ref{Mat-fer}) relies on the asymptotic factorisation of $\Pi_{mat}^{(n)}$ (as extensively discussed for scalars in \cite{BFPR2,BFPR3}). In fact, when $k$ rapidities $u_i$ and $v_i$ get shifted by a large quantity $\Lambda$, the matrix part factorises into the product of two matrix factors involving two disjoint proper subgroups of particles (up to a power of the shift):
\small
\be\label{FactPi}
\Pi_{mat}^{(n)}(\{u_{i=1}^{k} +\Lambda , u_{i=k+1}^n \},\{v_{i=1}^{k} +\Lambda , v_{i=k+1}^n  \}) \simeq \Lambda^{-4k(n-k)}\Pi_{mat}^{(k)}(\{u_{i=1}^k\},\{v_{i=1}^k\})\Pi_{mat}^{(n-k)}(\{u_{i=k+1}^n\},\{v_{i=k+1}^n\}) \,.
\ee
\normalsize
Formula (\ref{FactPi}) can be proven directly from the integral representation (\ref{Pi_mat^ff}). The leading contribution to $\Pi_{mat}^{(n)}$ in the limit $\Lambda\rightarrow\infty$ is obtained by re-absorbing the shift by $\Lambda $ on $k$ fermion and antifermion rapidities into the shift of $k$ integration variables $a,\,b,\,c$.\\
A proof of the polar structure (\ref{Mat-fer}) form factorisation can be sketched\footnote{A more detailed explanation can be found in \cite{BFPR3}.} as follows: if one shifts by $\Lambda\gg 1$ two fermion and two antifermion rapidities, say without loss of generality $u_1,\,u_2,\,v_1,\,v_2$, formula (\ref{FactPi}) for $k=2$ becomes
\small\be
\Pi_{mat}^{(n)}(u_1+\Lambda ,u_2+\Lambda, \{u_{i=3}^n\},v_1+\Lambda ,v_2+\Lambda, \{v_{i=3}^n\})
\simeq \Lambda^{-8(n-2)} \,\Pi_{mat}^{(2)}(u_1,u_2,v_1,v_2)
\Pi_{mat}^{(n-2)}(\{u_{i=3}^n\},\{v_{i=3}^n\}) \,.\nn
\ee\normalsize
The two-particle factor $\Pi_{mat}^{(2)}(u_1,u_2,v_1,v_2)$ enjoys the structure (\ref{Mat-fer}) and exhibits poles for $u_1-u_2=\pm i$,
$v_1-v_2=\pm i$ and $u_i-v_j=\pm 2i$. Since $\Pi_{mat}^{(n)}$ is invariant under permutations of the $u$'s and of the $v$'s, the same reasoning must hold for any arbitrary $4$-plet $\{u_i,u_j,v_k,v_l\}$. Then, structure (\ref{Mat-fer}) follows.
As a final remark, the factorisation of the polynomials $P^{(n)}$ is a straightforward consequence of (\ref{Mat-fer})
\be
P^{(n)}(\{u_{i=1}^{k} +\Lambda , u_{i=k+1}^n \},\{v_{i=1}^{k} +\Lambda , v_{i=k+1}^n  \}) \simeq \Lambda^{4k(n-k)}P^{(k)}(\{u_{i=1}^k\},\{v_{i=1}^k\})P^{(n-k)}(\{u_{i=k+1}^n\},\{v_{i=k+1}^n\}) \,. \nn
\ee

\section{The emergence of mesons}
\setcounter{equation}{0}
\label{ffcont}

The polar structure of the $SU(4)$ matrix factor (\ref{Pi_mat^ff}) and the properties of the polynomials $P^{(n)}$ (\ref{Mat-fer}) play a crucial role to unravel how fermions and antifermions coalesce into bound states, hereafter dubbed as `mesons' \cite{BSV3,BFPR1}. From the standpoint of the Bethe Ansatz equations, mesons do not participate in the particle spectrum at finite coupling, as they lie outside of the physical sheet \cite{BSV3}: on the contrary, they come into existence for infinitely large values of the coupling, when they provide a dominant contribution to the OPE, in contrast to unpaired fermions and antifermions. In this section we complete the work initiated in \cite{BFPR1}, where only two couples fermion--antifermion were considered, by generalising up to any number of particles. Mesons in turn bind up to form further composite states, as it will be elucidated in Section \ref{Mes-Ins}.

\begin{figure} [htbp]
\centering
\includegraphics[width=0.85\textwidth]{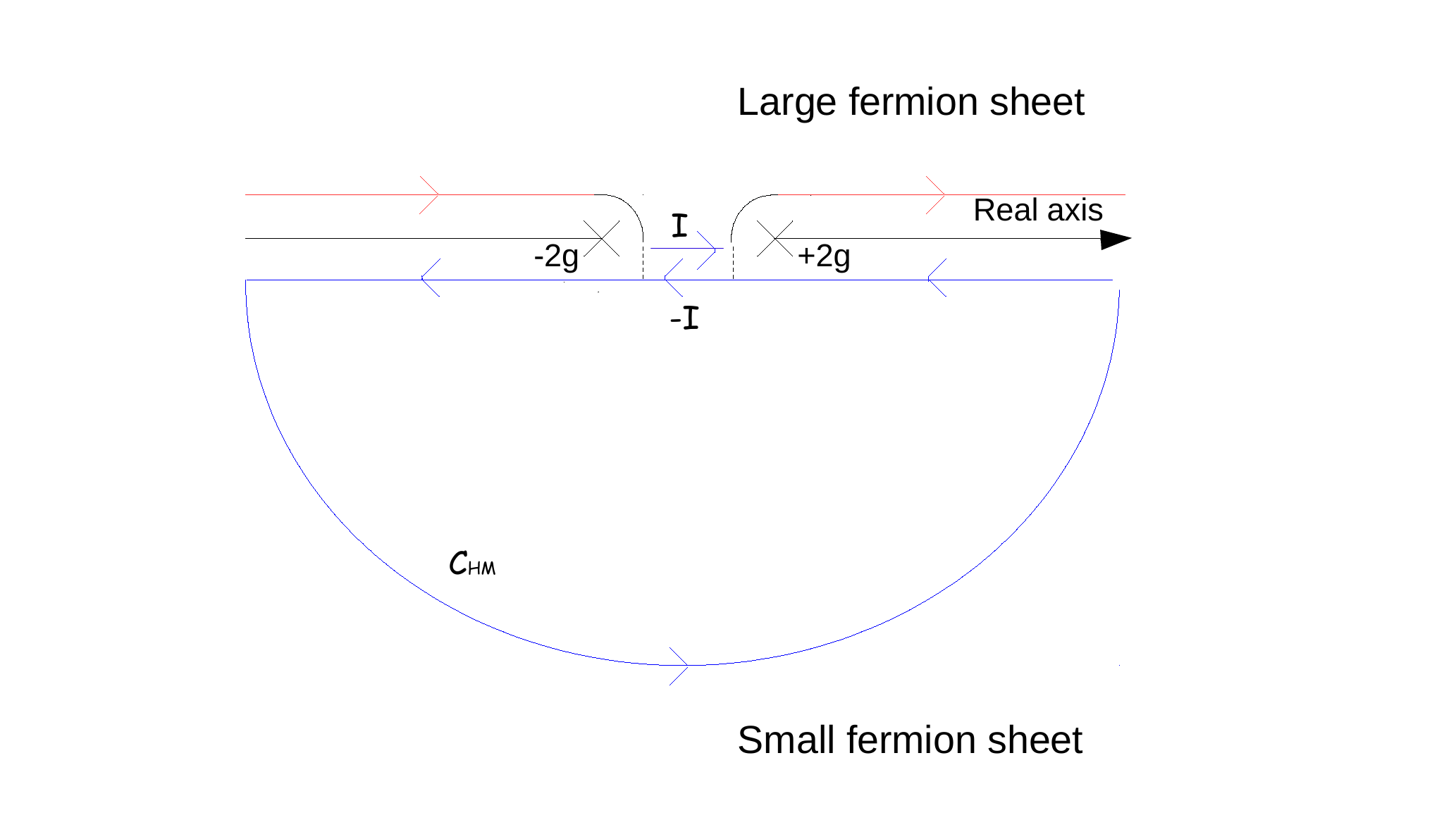}
\caption{Adding and subtracting the open interval $\m{I}=[-2g,+2g]$ depicted in figure, the integration contour $\m{C}_{S}$ can be seen as a sum of a closed curve $\m{C}_{HM}$, with half-moon shape, and an interval $\m{I}$ in the small sheet just below the branch cut (which involves unphysical values of the rapidity).}
\label{Cfigura2}
\end{figure}

To ease our task, we reformulate (\ref{WLferm}) into
\be\label{Meson}
W^{(n)}_f =\frac{1}{n!}\int_{\m{C}_S}\displaystyle\prod_{i=1}^n\frac{du_i}{2\pi}I_n(u_1,\ldots ,u_n)\displaystyle\prod_{i<j}^np(u_{ij}) \ ,
\ee
where we highlighted the `short-range' (meson-meson) potential
\be\label{pol-pot}
p(u_{ij})\equiv \displaystyle\frac{u_{ij}^2}{u_{ij}^2+1} \, , \quad u_{ij}=u_i-u_j \, ,
\ee
whose meaning will be clarified in next section. In (\ref{Meson}) we enclosed the integrals on the antifermionic rapidities $v_j$ inside the functions
\small
\be\label{I_n}
I_n(u_1, \ldots , u_n)\equiv\frac{1}{n!}\int_{\m{C}_S}\displaystyle\prod_{i=1}^n\frac{dv_i}{2\pi} R_n(u_1,\ldots ,u_n,v_1,\ldots ,v_n)P^{(n)}(u_1,\ldots ,u_n,v_1,\ldots ,v_n)
\displaystyle\prod_{i,j=1}^n h(u_i-v_j)\displaystyle\prod_{i<j}^n p(v_{ij}) \ ,
\ee
\normalsize
where we defined the short-range (fermion-antifermion) potential
\be
h(u_i-v_j)=\frac{1}{(u_i-v_j)^2+4} \, .
\ee
The regular part, with no poles nor zeroes in the rapidities $u_i,\,v_i\,$, is encoded in the function $R_n$, which is related to the dynamical factor (\ref{dyn}) via the definition
\be\label{R_n}
R_n(u_1,\ldots ,u_n,v_1,\ldots ,v_n)\displaystyle\prod_{i<j}^n u_{ij}^2v_{ij}^2 \equiv
\Pi_{dyn}^{(n)}(u_1,\ldots ,u_n,v_1,\ldots ,v_n)\displaystyle\prod_{i=1}^n\hat{\mu}_f(u_i)\hat{\mu}_f(v_i) \ ,
\ee
also involving the measure and the propagation phase, combined into
\be
\hat{\mu}_f(u)\equiv \mu_f(u)e^{-\tau E_f(u) + i\sigma p_f(u)} \ .
\ee
The integration in (\ref{I_n}) has been safely restricted from the curve $\m{C}$ to its small-fermion sheet section $\m{C}_S$ ($cf.$ Figure \ref{Csmall}): in turn, $\m{C}_S$ can be thought as the superposition of a closed contour $\m{C}_{HM}$, entirely lying in the lower half plane and oriented in the counterclockwise direction, plus the segment $\mathcal{I}=[-2g,+2g]$, oppositely oriented \cite{BFPR1}, as portrayed in Figure \ref{Cfigura2}.
The choice of the contour $\m{C}_S$ does not rely on a specific value of the coupling constant $g$, nevertheless as far as the strong coupling regime is concerned, we benefit from a crucial simplification. Indeed, if we decompose the function $I_n$ into the sum
$$I_n=I_n^{closed}+I_n^r$$
of the term $I_n^{closed}$, corresponding to the expression (\ref{I_n}) with the contour $\m{C}_S$ replaced by $\m{C}_{HM}$, plus the remainder $I_n^r$, the latter contribution turns out to be subdominant \cite{BFPR1}.\\
To exemplify the approach to the integrations in (\ref{I_n}), we first consider the simplest non-trivial case, $n=2$.\\
\ \medskip\ \\
\textbf{$\bullet\ n=2\,$:}\\
When evaluating the integrals
\be\label{I_2}
I_2^{closed}(u_1,u_2)=\frac{1}{2}\int_{\m{C}_{HM}}\frac{dv_1 dv_2}{(2\pi)^2}\frac{R_2(u_1,u_2,v_1,v_2)P^{(2)}(u_1,u_2,v_1,v_2)}{\left[(u_1-v_1)^2+4\right]\left[(u_1-v_2)^2+4\right]\left[(u_2-v_1)^2+4\right]\left[(u_2-v_2)^2+4\right]}\frac{v_{12}^2}{v_{12}^2+1}
\ee
by residues, one observes that the poles arrange themselves in strings in the complex plane, with real coordinates $u_j$: indeed, if the first pole to be considered is $v_i=u_j-2i$\,, any further residue around the same real rapidity is placed below the previous one at a distance $-i$\,, $i.e.$ the general form $v_i=u_j-(2+\kappa)i$ (with $\kappa =0,1,\dots $) is found. Sticking on (\ref{I_2}), three independent residue configurations occur, namely $(u_1-2i,u_2-2i)$, $(u_1-2i,u_1-3i)$ and $(u_2-2i,u_2-3i)$, each one with a multiplicity $2!$ owing to the symmetry of $I_2^{closed}$ under permutations of $v_i$: these configurations are respectively denoted in the following as $(1,1)$, $(2,0)$ and $(0,2)$, for compactness. The explicit form of $P^{(2)}$ entails that $(2,0)$ and $(0,2)$ actually give no contribution, since
\be\label{nullYoung}
P^{(2)}(u_1,u_2,u_1-2i,u_1-3i)=0 \ ,
\ee
while the only configuration that matters is $(1,1)$, for which $P^{(2)}$ takes the simple form
\be\label{Young-2}
P^{(2)}(u_1,u_2,u_1-2i,u_2-2i)=16\left[(u_{12}^2+16)(u_{12}^2+1)\right] \ :
\ee
as a result,
\be
I_2^{closed}(u_1,u_2) = R_2(u_1,u_2, u_1-2i,u_2-2i) \ ,
\ee
which on a physical ground hints that fermions and antifermions (whose rapidities differ by $2i$) form bound states.
\ \medskip\ \\
\textbf{$\bullet$ Arbitrary $n$:}\\
As for $n=2$, also in the general case with arbitrary $n$ the only residue configuration that gives a contribution to $I_n$ turns out to be $(1,1,\ldots,1)$, $i.e.$ only poles of the kind $v_i=u_j-2i$ are involved. In fact, the strict constraint
\be
P^{(n)}(u_1,\ldots ,u_n, u_1-2i, u_1-3i , v_3, \ldots , v_n)=0 \,,
\ee
resulting from (\ref{RecConj}), rules out all the residue configurations except for $(1,1,\ldots,1)$, so that one can assert, quite formally,
\be\label{P(Y)=0}
P^{(n)}(Y\neq (1,1\ldots ,1,1))=0 \,.
\ee
Taking into account the proper combinatorial factor, the contribution arising from the aforementioned configuration reads
\be\label{1111}
(1,1,\ldots ,1,1)=\frac{(-1)^n}{4^n}\frac{P^{(n)}(u_1,\ldots ,u_n,u_1-2i,\ldots ,u_n-2i)R_n(u_1,\cdots ,u_n,u_1-2i,\ldots ,u_n-2i)}{\displaystyle\prod_{i<j}^n(u_{ij}^2+16)(u_{ij}^2+1)}
\ee
where $P^{(n)}$ can be given an explicit form (\ref{fundP})
\be\label{Pconj}
P^{(n)}(u_1,\ldots ,u_n,u_1-2i,\ldots ,u_n-2i)=4^n\displaystyle\prod_{i<j}^n(u_{ij}^2+16)(u_{ij}^2+1)
\ee
as shown in Appendix \ref{appA}, allowing us to retrieve for (\ref{I_n}) an expression
\be\label{I_n-clo}
I_n^{closed}(u_1,\ldots ,u_n)=(-1)^n R_n(u_1,\ldots ,u_n, u_1-2i,\ldots ,u_n-2i)
\ee
which highlights how fermion and antifermion rapidities pair up to form complex two-string, with spacing $2i$. A comparison with (\ref{dyn}), (\ref{R_n}) suggests to interpret these two-strings as bound states \cite{BSV3,BFPR1}, dubbed mesons, whose energy is the sum of energies of the single components, as well as their momentum,
\be
E_M(u)\equiv E_f(u+i)+E_f(u-i), \quad p_M(u)\equiv p_f(u+i)+p_f(u-i) \ .
\ee
The meson pentagon transition amplitude can be recognised in the expression
\be
P^{(MM)}(u|v) = -(u-v)(u-v+i) P^{(ff)}(u+i|v+i)P^{(ff)}(u-i|v-i)|P^{(f\bar f)}(u+i|v-i)P^{(f\bar f)}(u-i|v+i) \ ,
\ee
although, for later purposes, it is worth to introduce the regular function $P^{(MM)}_{reg}$, without poles nor zeroes, related to $P^{(MM)}$ via
\be
P^{(MM)}(u|v)
= \frac{u-v+i}{u-v}\,P^{(MM)}_{reg}(u|v) \ .
\ee
Accordingly, the (hatted) measure can be coherently traced in the formula
\be\label{measureM}
\hat{\mu}_M(u) \equiv \mu_M(u)e^{-\tau E_M(u) + i\sigma p_M(u)}
= -\frac{\hat{\mu}_f(u+i)\hat{\mu}_f(u-i)}{P^{(f\bar f)}(u+i|u-i)P^{(f\bar f)}(u-i|u+i)} \ ,
\ee
both from a direct inspection to (\ref{R_n}) or from its relation to the pentagonal amplitude, mimicking (\ref{misura_fermione})
\be
\mbox{Res}\,_{v=u}\,P^{(MM)}(u|v)=\frac{i}{\mu_M(u)} \ .
\ee
These identifications lead us to recast (\ref{I_n-clo}) into the expression
\be\label{conj}
I_n^{closed}(u_1,\ldots ,u_n) 
 = \frac{\displaystyle\prod_{i=1}^n\hat{\mu}_M(u_i-i)}{\displaystyle\prod_{i<j}^n P^{(MM)}_{reg}(u_i-i|u_j-i)P^{(MM)}_{reg}(u_j-i|u_i-i)} \ ,
\ee
that, once plugged into (\ref{Meson}), legitimises us to reformulate the fermion contribution to the hexagon Wilson loop (\ref{WL}) in terms of these novel bound states, into the series
\ba\label{SingMes}
W_f &=& W_{M}+\dots = \\
&=& \sum_{n=0}^{\infty}\frac{1}{n!}\int_{\m{C}_S}\displaystyle\prod_{i=1}^n\frac{du_i}{2\pi}\hat{\mu}_M(u_i-i)\displaystyle\prod_{i<j}^n\frac{1}{P^{(MM)}_{reg}(u_i-i|u_j-i)
P^{(MM)}_{reg}(u_j-i|u_i-i)}\displaystyle\prod_{i<j}^n p(u_{ij}) +\dots \nn
\ea
where the dots are to remind that some terms, coming from the integrations on the interval $\mathcal{I}$ and originally included in $I_n$, were discarded\footnote{In order to fully reconstruct the fermionic contribution to the Wl at finite coupling, the integrations performed along the contours $C_L$ shall be added too.} while considering the leading contribution $I_n^{closed}$.

As a conclusive remark, formula (\ref{SingMes}) means that in the large coupling regime, unpaired fermions and antifermions give the way to the formation of mesons, nevertheless it still makes sense even at finite coupling: indeed one can still recognise a contribution ascribable to these effective particles, and associate them (at least formally) to a pentagon amplitude and a measure.

\section{Resummation: from the OPE to the TBA}
\setcounter{equation}{0}
\label{Mes-Ins}

The peculiar form of the short range potential (\ref{pol-pot}) allows us to trade, at leading order, the sum on mesons in $W_M$ (\ref{SingMes}) for a sum performed on `TBA effective bound states': we talk about `effective bound states', in that they are not associated to any new node in the Y-system for scattering amplitudes \cite{YSA}.
As a matter of fact, the formula (\ref{SingMes}) for $W_M$ shares its form with the instanton partition function $\mathcal{Z}$ of $\mathcal{N}=2$ theories \cite{Nek}, and from this perspective the large coupling $g\sim 1/\epsilon_2$ for $W_M$ corresponds to the so-called Nekrasov-Shatashvili (NS) limit of $\mathcal{Z}$, where the $\Omega$-background $\epsilon_2$ approaches zero \cite{NekSha}. Taking advantage of this analogy, we analyse the strong coupling limit of $W_M$, write $W_M$ as an expectation value of a Fredholm determinant, to finally reconstruct the full set of TBA equations for the hexagon \cite{YSA, Anope} straight from the OPE series of fermions (and at a subsequent stage including gluons). In comparison with \cite{FPR}, the element of novelty is that the existence of mesons is not \textit{assumed} as a hypothesis, but rather derived.

We point out that the technique we are about to outline constitutes a powerful tool to step over the leading order in $1/g$ and eventually obtain information at finite coupling: see for instance \cite {BouFio1, BouFio2} where the subleading corrections are computed for the Nekrasov function $\mathcal{Z}$. The method will be discussed more thoroughly in Appendix \ref{NekApp}, where it will be applied to $\mathcal{Z}$.

\subsection{Mesons and the $\mathcal{N}=2$ Nekrasov function}

To begin with, we introduce the Nekrasov function $\mathcal{Z}$, which accounts for the instanton effects in some $\mathcal{N}=2$ gauge theories \cite{Nek}. The instanton contribution to the total partition function is computed in a deformed space-time, the so-called $\Omega$-background, which depends on two parameters $\epsilon_1$ and $\epsilon_2$. The (instanton part of) Seiberg-Witten prepotential of the original theory is obtained by sending to zero the deformation parameters in the combination $\epsilon_1 \epsilon_2 \log\mathcal{Z}$.

As far as the connection with $W_M$ is concerned, we are interested only in the dependence on one\footnote{They appear symmetrically in the definition of $\mathcal{Z}$.} of the two parameters, which we call simply $\epsilon$, leaving the other concealed in the various functions appearing in the definition of $\mathcal{Z}$. The Nekrasov function enjoys the following integral representation
\be\label{Z0}
\mathcal{Z}=\sum_{n=0}^{\infty}\frac{q^n}{n!\epsilon ^n}\int\displaystyle\prod_{i=1}^n\frac{du_i}{2\pi i}Q(u_i)\displaystyle\prod_{i<j}^n e^{\epsilon G(u_{ij})}\displaystyle\prod_{i<j}^n\frac{u_{ij}^2}{u_{ij}^2-\epsilon^2} \,,
\ee
where the sum is performed over the instanton number $n$, while the integrals are over the instanton coordinates $u_i$, resembling the grand canonical partition function for an interacting classical one dimensional gas. Here $q$ is the instanton parameter related to the complex coupling of the gauge theory, $Q(u)$ is a rational function depending on the gauge group and the matter content. The interaction kernels $G(u)$ and $\displaystyle\prod_{i<j}^n\frac{u_{ij}^2}{u_{ij}^2-\epsilon^2}$ are universal for $SU(N_c)$ gauge groups and represent, respectively, a sort of long range and short-range two-body potential acting between the instantons. This distinction is meaningful in the NS limit $\epsilon\to 0$, where the long-range potential is smooth and the short-range becomes singular. We stress that the functions $Q$ and $G$ depend also on $\epsilon$ but they enjoy a smooth limit for $\epsilon\to 0$, so that in the following we do not consider this dependence.

The analogy between the meson series $W_M$ (\ref{SingMes}) and $\mathcal{Z}$ (\ref{Z0}), as anticipated in \cite{BFPR1}, is made manifest upon identifying $\epsilon\sim \frac{i}{g}$, so that the strong coupling limit for the Wilson loop corresponds to the NS limit for $\mathcal{Z}$. Indeed, in this regime the main contribution comes from the region where the variables $u_i$ are very large, thus the rescaled rapidities $u_i\equiv 2g \bar{u}_i$ are suitably defined. The polar part reads then
\be\label{Wshort}
p(u_{ij})=\frac{u_{ij}^2}{u_{ij}^2+1} = \frac{\bar{u}_{ij}^2}{\bar{u}_{ij}^2+\frac{1}{4g^2}} \, ,
\ee
making manifest its interpretation as a short-range interaction, $cf.$ (\ref{Z0}).
On the other hand, the role of the long-range interaction $e^{\epsilon G(u_{ij})}$ is taken by
\be\label{Wlong}
\frac{1}{P^{(MM)}_{reg}(u_i-i|u_j-i)
P^{(MM)}_{reg}(u_j-i|u_i-i)}\simeq 1-\frac{2\pi}{\sqrt{\lambda}}K_M(\theta_i,\theta_j) \,,
\ee
where the kernel $K_M$ \cite{FPR, BB1} is conveniently expressed in terms of the hyperbolic rapidity $\bar{u}\equiv \coth\theta$:
\be
K_M(\theta_i,\theta_j)=-\frac{\sinh(2\theta_i)\sinh(2\theta_j)}{\cosh(\theta_i-\theta_j)} \,.
\ee
When it comes to the differences between the series (\ref{Z0}) and (\ref{SingMes}), it shall be pointed out that in $W_M$ the long-range potential (\ref{Wlong}) and the measure (\ref{measureM}) are free of poles in the small fermion sheet, while their counterparts in $\mathcal{Z}$, namely $G(u)$ and $Q(u)$, exhibit poles inside the integration region. Furthermore, since the integration contours in $\mathcal{Z}$ (\ref{Z0}) are closed,  to make the analogy more stringent a segment $\mathcal{I}$ (whose contribution to the integral is suppressed in the strong coupling limit \cite{BFPR1}) shall be added to the open path $\m{C}_S$, building up $\m{C}_{HM}$, $cf.$ Figure \ref{Cfigura2}. These differences though do not affect the mechanism responsible for the formation of the bound states, as it is driven both in (\ref{Z0}) and (\ref{SingMes}) by the poles of the short range interaction.

\subsection{Mesons, bound states and TBA}\label{MesTBA}

At strong coupling, mesons coalescence into bound states: we will follow two alternative approaches to describe the process, first via a Mayer expansion, then again through a Hubbard-Stratonovich transformation and the Cauchy identity, reprising \cite{proc}.

\vspace{0.5cm}
\noindent\textbf{$\bullet $\ Mayer expansion}\\
Recalling \cite{Bou,MenYang}, it is fruitful to expand the polar part in clusters
\be\label{clusters}
\displaystyle\prod_{i<j}^n\frac{u_{ij}^2}{u_{ij}^2+1}=\displaystyle\prod_{i<j}^n\left[1-\frac{1}{u_{ij}^2+1}\right]=\sum_{C_n}\displaystyle\prod_{(i,j)\in C_n}\left[-\frac{1}{u_{ij}^2+1}\right] \ :
\ee
upon plugging into (\ref{SingMes}), each connected sub-cluster in the sum (\ref{clusters}) gets associated to a bound state.\footnote{A specific configuration of bound states can be built in multiple ways, and all of these shall be summed up.} More precisely, if we have $l$ connected sub-clusters with $n_i$ particles each, satisfying $n_1+\cdots n_l=n$, we integrate out $n_i-1$ variables for each cluster: the remaining $l$ unintegrated variables can be assimilated to the coordinates of the bound states.
With the aid of (\ref{In}), we get:
\be\label{int-dil}
\int_{\m{C}_{HM}}\displaystyle\prod_{i=1}^{a-1}\frac{du_i}{2\pi}\sum_{C^c_a}\displaystyle\prod_{(i,j)\in C^c_a}\left[-\frac{1}{u_{ij}^2+1}\right]=\int_{\m{C}_{HM}}\displaystyle\prod_{i=1}^{a-1}\frac{du_i}{2\pi}\displaystyle\prod_{i<j}^a\frac{u_{ij}^2}{u_{ij}^2+1}
=\frac{a!}{a^2} \,.
\ee
The multiple integral (\ref{int-dil}) allows us to write $W_M$ (\ref{SingMes}) at strong coupling as a sum over bound states of mesons, $W_M=W_{M,bound}+\cdots$
\be\label{BoundMes}
W_{M,bound}=\sum_{N=0}^{\infty}\frac{1}{N!}\sum_{a_1=1}^{\infty}\cdots\sum_{a_N=1}^{\infty}\int_{\m{C}_S}\displaystyle\prod_{i=1}^N\frac{du_i}{2\pi}\hat{\mu}_{M,a_i}(u_i)\displaystyle\prod_{i<j}^N\frac{1}{P^{(MM)}_{a_i,a_j}(u_i|u_j)P^{(MM)}_{a_j,a_i}(u_j|u_i)} \,,
\ee
where the hatted measure of a bound state and the pentagonal transition at strong coupling are given \cite{FPR} by
\ba
\hat{\mu}_{M,a}(u) &\simeq & \frac{[\hat{\mu}_M(u)]^a}{a^2} \\
P^{(MM)}_{a,b}(u|v) &\simeq & \left[P^{(MM)}_{reg}(u|v)\right]^{ab} \,.
\ea
In (\ref{BoundMes}), $N$ represents the number of particles (single or bound mesons), and $a_i$ is the number of fundamental constituents of the $i$-th particle.
\vspace{0.5cm}

\noindent\textbf{$\bullet $\ Fredholm determinant $+$ Path integral}\\
A more rigorous and elegant proof of the equivalence between the series (\ref{SingMes}) and (\ref{BoundMes}) relies on a Hubbard-Stratonovich transformation and the Cauchy identity, crucial to formulate $W_M$ in terms of a Fredholm determinant. We associate a boson field $X^M(u)$ to mesons, through its propagator
\be\label{bosonM}
e^{\left\langle X^M(u_i)X^M(u_j)\right\rangle} \equiv \frac{1}{P^{(MM)}_{reg}(u_i-i|u_j-i)
P^{(MM)}_{reg}(u_j-i|u_i-i)}\equiv e^{G_M(u_i,u_j)} \ :
\ee
by performing the functional Gaussian integrations, one obtains the renowned identity
\be\label{gauss}
\prod_{i<j}^n e^{\langle X^M(u_i)X^M(u_j)\rangle}=\displaystyle\prod_{i=1}^n e^{-\frac{1}{2}G_M(u_i,u_i)}\langle \prod_{k=1}^n e^{X^M(u_k)}  \rangle \,,
\ee
which allows writing the Wl as
\be
W_M \simeq \left\langle \sum_{n=0}^{\infty}\frac{1}{n!}\int_{\m{C}_S}\displaystyle\prod_{i=1}^n\frac{du_i}{2\pi }\hat{\mu}_M(u_i-i)e^{X^M(u_i)}\displaystyle\prod_{i<j}^n p(u_{ij})\right\rangle \,.
\ee
Above, the diagonal term from the Gaussian identity (\ref{gauss}), corresponding to the propagator evaluated in $u_i=u_j$, was neglected, for it is subleading in the $1/g\rightarrow 0$ limit.
We recall the Cauchy identity
\be
\displaystyle\prod_{i<j}^n p(u_{ij})=\frac{1}{i^n}\det\left(\frac{1}{u_i-u_j-i}\right)
\ee
and define the matrix $M$
\be
M(u_i,u_j)\equiv \frac{\left[\hat{\mu}_M(u_i-i)e^{X^M(u_i)}\hat{\mu}_M(u_j-i)e^{X^M(u_j)}\right]^{1/2}}{u_i-u_j-i} \,,
\ee
in order to find a compact representation for the Wl at strong coupling
\be\label{WFre}
W_M \simeq \left\langle\det\left(1 + M\right)\right\rangle = \left\langle \exp\left[\sum_{n=1}^{\infty}\frac{(-1)^{n+1}}{n}\textit{Tr}(M^n)\right]\right\rangle \,.
\ee
As customary, the trace is defined as
\be
\textit{Tr}(M^n) \equiv \int_{\m{C}_s}\displaystyle\prod_{i=1}^n\frac{du_i}{2\pi i}\hat{\mu}_M(u_i-i)e^{X^M(u_i)}\displaystyle\prod_{i=1}^n\frac{1}{u_i-u_{i+1}-i}, \quad u_{n+1}\equiv u_1 \,,
\ee
whose leading order is obtained by evaluating the residues of poles enclosed inside $\m{C}_{HM}=\m{C}_{s} \cap \m{I}$
\be\label{TrLeading}
\textit{Tr}(M^n) \simeq \frac{(-1)^{n-1}}{n}\int_{\m{C}_s}\frac{du}{2\pi}\hat{\mu}^n_M(u-i)e^{nX^M(u)}\simeq \frac{(-1)^{n-1}}{n}\int_{\m{C}_s}\frac{du}{2\pi}\hat{\mu}^n_M(u)e^{nX^M(u)}
\ee
where the shifts have been neglected, for the rapidities get rescaled $u=2g\bar u$.
Taking into account that in the large $g$ limit $\mu_M(u)\simeq -1$, the Wl can be recast as \cite{FPR}
\be\label{WLi2}
W_M\simeq\left\langle\exp\left[-\int_{\m{C}_s}\frac{du}{2\pi}\mu_M(u) Li_2\left[-e^{-\tau E_M(u)+i\sigma p_M(u)} e^{X^M(u)}\right]\right]\right\rangle \,.
\ee
Formula (\ref{WLi2}) suggests a reformulation in terms of a path integral
\be
W_M\simeq \int DX^M e^{-S[X^M]} \,,
\ee
where the action reads
\be\label{actionM}
S[X^M]=-\frac{1}{2}\frac{\sqrt{\lambda}}{2\pi}\int d\theta d\theta' X^M(\theta)K^{-1}_M(\theta,\theta')X^M(\theta')+\int\frac{d\theta}{2\pi}\mu_M(\theta)Li_2\left[-e^{-\sqrt{2}E(\theta)}e^{X^M(\theta)}\right]
\ee
and the inverse kernel is defined according to
\be
\int d\theta' K_M(\theta,\theta')K_M^{-1}(\theta',\theta'')=\delta (\theta-\theta'') \,.
\ee
The cross ratios are incorporated in the function $E(\theta)\equiv \sqrt{2}\tau \cosh\theta -i\sqrt{2}\sigma\sinh\theta$ and, in terms of the hyperbolic rapidity $\q$, the measure reads
\be
\mu_M(\theta)=\frac{\sqrt{\lambda}}{2\pi}\frac{2}{\sinh^2\theta} \,.
\ee
Finally, since the action is proportional to the coupling $\sqrt{\lambda}$, we can perform a saddle point approximation which leads to the central node equation of the TBA for scattering amplitudes \cite{YSA, Anope}
\be\label{TBAamp}
X^M(\theta) - \int\frac{d\theta'}{2\pi}\mu_M(\theta')G_M(\theta,\theta')\log\left[1+e^{-\sqrt{2}E(\theta)}e^{X^M(\theta)}\right]=0 \,.
\ee
The leading order of the logarithm of $W_M$, then, approximates the critical action, $i.e.$ the action (\ref{actionM}) evaluated through the solution of (\ref{TBAamp}):
\be\label{YYamp}
-\ln W_M\simeq S_c = \frac{1}{2}\int d\theta \mu_M(\theta)X^M(\theta)\log\left[1+e^{-\sqrt{2}E(\theta)}e^{X^M(\theta)}
\right] +\int\frac{d\theta}{2\pi}\mu_M(\theta)Li_2\left[-e^{-\sqrt{2}E(\theta)}e^{X^M(\theta)}\right] \,.
\ee

\subsection{Gluons and fermions at strong coupling}\label{glufer}

In order to fully understand the behaviour of mesons, their interaction with gluons needs to be taken into account. We consider the contribution $W_{f,g}$ from intermediate states including $n$ fermions with rapidities $u_i$, $n$ anti-fermions with rapidities $v_i$ and $m$ bound states of gluons, each characterised by rapidity $u_{g,i}$ and helicity $a_i=\pm 1,\pm 2,\pm 3,\dots$ (such that $|a_i|$ also denotes the number of bound gluons):
\ba\label{W1}
&& W_{f,g} =\sum _{m,n=0}^{+\infty} \frac{1}{m!n!}\sum _{a_1}\ldots \sum _{a_m}
\int \prod _{k=1}^m \frac{du_{g,k}}{2\pi} \hat\mu_{g,a_k}(u_{g,k})\,e^{ia_k\phi} \prod _{i<j=1}^m\frac{1}{P^{(gg)}_{a_ia_j}(u_{g,i}|u_{g,j})P^{(gg)}_{a_ja_i}(u_{g,j}|u_{g,i})} \cdot  \nn\\
&& \cdot \int \prod _{i=1}^n \frac{du_i}{2\pi} I^g_n(u_1,...,u_n, u_{g,1},...,u_{g.m}) \prod _{i<j=1}^n \frac{u_{ij}^2}{u_{ij}^2+1} \prod _{i=1}^m \prod _{j=1}^n \frac{1}{P_{a_i}^{(gf)}(u_{g,i}|u_j)P_{a_i}^{(fg)}(u_j|u_{g,i})} \, ,
\ea
where the function $I_n$ given in (\ref{I_n}) is generalised so to include the interaction between gluons and anti-fermions
\be\label{Ign}
I^g_n(u_1,...,u_n, u_{g,1},...,u_{g,m})\equiv \frac{1}{n!}\int_{C}\displaystyle\prod_{i=1}^n\frac{dv_i}{2\pi} \frac{R_n(\{u_i\},\{v_j\})P^{(n)}(\{u_i\},\{v_j\})}{\displaystyle\prod _{i=1}^m \prod _{j=1}^n P_{a_i,\epsilon _i}^{(g\bar f)}(u_{g,i}|v_j)P_{a_i,\epsilon _i}^{(\bar fg)}(v_j|u_{g,i})} \displaystyle\prod_{i,j=1}^n h(u_i-v_j)
\displaystyle\prod_{i<j}^n p(v_{ij}) \,,
\ee
while the hatted measure is defined as
\be
\hat\mu_{g,a}(u)\equiv \mu_{g,a}(u)\,e^{-\tau E_{g,a}(u)+i\sigma p_{g,a}(u)} \,.
\ee
The evaluation by residues of the integrals on the antifermionic rapidities in (\ref {Ign}) produces the fusion of fermion-gluon amplitudes into analogous mesonic quantities
\ba
P_{a}^{(gM)}(u_g|u) &=& P_{a}^{(gf)}(u_g|u+i)P_{a}^{(g\bar f)}(u_g|u-i) \label {Pmg} \\
P_{a}^{(Mg)}(u|u_g) &=& P_{a}^{(fg)}(u+i|u_g)P_{a}^{(\bar fg)}(u-i|u_g) \ ; \nn
\ea
for explicit expressions, the reader is addressed to Appendix \ref{use_f}. As a by-product, unbound fermions are wiped away from (\ref{W1}):
\ba\label  {W2}
W_{f,g} &\simeq & W_{M,g} = \sum _{m,n=0}^{+\infty} \frac{1}{m!n!}\sum _{a_1}\ldots \sum_{a_m}
\int\prod_{k=1}^m \frac{du_{g,k}}{2\pi} \hat\mu_{g,a_k}(u_{g,k})\,e^{ia_k\phi}\prod _{i=1}^n \frac{du_i}{2\pi} \hat \mu_M(u_i-i) \cdot \nn\\
&\cdot &\prod_{i<j=1}^n \frac{u_{ij}^2}{u_{ij}^2+1}\prod _{i<j=1}^n \frac{1}{P^{(MM)}_{reg}(u_i-i|u_j-i)P^{(MM)}_{reg}(u_j-i|u_i-i)} \cdot \\
&\cdot & \prod_{i<j=1}^m\frac{1}{P^{(gg)}_{a_ia_j}(u_{g,i}|u_{g,j})P^{(gg)}_{a_ja_i}(u_{g,j}|u_{g,i})}
\prod _{i=1}^m \prod _{j=1}^n \frac{1}{P_{a_i}^{(gM)}(u_{g,i}|u_j-i)P_{a_i}^{(Mg)}(u_j-i|u_{g,i})} \,.\nn
\ea
Mimicking (\ref{bosonM}), boson fields may be associated to bound states of gluons through their propagators
\ba
\exp \langle X^{g}_{a}(u) X^{g}_b(v) \rangle &=& \frac{1}{P^{(gg)}_{ab}(u|v)P^{(gg)}_{ba}(v|u)} \\
\exp \langle X^{g}_{a}(u) X^{M}(v) \rangle &=& \frac{1}{P_{a}^{(gM)}(u|v-i)P_{a}^{(Mg)}(v-i|u)} \,.\nn
\ea
The remarkable simplifications occurring at strong coupling \cite{FPR}
\ba
&& P^{(gg)}_{ab}(u|v)P^{(gg)}_{ba}(v|u)\simeq \left[P^{(gg)}(u|v)P^{(gg)}(v|u)\right]^{|ab|} \quad \quad P_{a}^{(gM)}(u|v)\simeq \left[P^{(gM)}(u|v)\right]^{|a|} \nn \\
&& \hat\mu_{g,a}(u)\simeq \frac{[\hat\mu_g(u)]^{|a|}}{a^2} \quad \quad \hat \mu _g(u)=-e^{-\tau E_g(u) +i\sigma p_g(u)} \nn
\ea
suggest that we can set $X^{g}_a(u)=|a| X^{g}(u)$ to rearrange $W_{M,g}$ as:
\ba\label{WMg}
W_{M,g} &=&\sum _{m,n=0}^{+\infty} \frac{1}{m!n!}\sum _{a_1}\ldots \sum _{a_m}
\int \prod _{k=1}^m \frac{du_{g,k}}{2\pi} \frac{[\hat \mu_{g}(u_{g,k})]^{|a_k|} \,e^{ia_k\phi}}{a_k^2} \prod _{i=1}^n \frac{du_i}{2\pi} \hat \mu_M(u_i-i) \det _{i,j} \frac{1}{u_{ij}-i} \cdot \nn \\
&\cdot & \langle e^{X^M(u_1)}...e^{X^M(u_n)} e^{|a_1| X^g(u_{g,1})}...e^{|a_m| X^g(u_{g,m})} \rangle = \nn \\
&=& \left\langle \sum _{m=0}^{+\infty}\frac{1}{m!}\sum_{a_1}\ldots\sum_{a_m}\int\prod_{k=1}^m\frac{du_{g,k}}{2\pi}
\frac{[\hat\mu_{g}(u_{g,k})]^{|a_k|}\,e^{ia_k\phi}}{a_k^2} e^{|a_k| X^g(u_{g,k})} \right. \cdot \nn \\
&\cdot & \left. \sum _{n=0}^{+\infty} \frac{1}{n!} \int \prod _{i=1}^n \frac{du_i}{2\pi} \hat \mu_M(u_i-i) \det _{i,j} \frac{1}{u_{ij}-i}  e^{X^M(u_i)} \right \rangle \,.
\ea
The latter sum already appeared in Section \ref{MesTBA}, giving as outcome (\ref{WLi2}), whereas the former amounts to
\ba
&& \sum _{m=0}^{+\infty} \frac{1}{m!}\left[ \left(\sum_{a=1}^{+\infty}+\sum_{a=-\infty}^{-1}\right)\int \frac{du}{2\pi}
\frac{[\hat\mu_{g}(u)]^{|a|}}{a^2}\,e^{ia\phi}\,e^{|a|X^{g}(u)} \right]^m  = \\
&& = \exp \Bigl[ \int \frac{du}{2\pi} \textrm{Li}_2 \left( -e^{-\tau E_{g}(u)+i\sigma p_{g}(u)+i\phi} e^{X^{g}(u)} \right) +  \int \frac{du}{2\pi} \textrm{Li}_2 \left ( -e^{-\tau E_{g}(u)+i\sigma p_{g}(u)-i\phi}
e^{X^{g}(u)} \right ) \Bigr] \,. \nn
\ea
Eventually (\ref{WMg}) takes the form \cite{FPR} we need to reconstruct the full set of TBA equations for the hexagonal Wilson loop:
\ba
&& W_{M,g} =\Bigl \langle
\exp \int \frac{du}{2\pi} \textrm{Li}_2 \left ( -e^{-\tau E_{M}(u)+i\sigma p_{M}(u)}
e^{X^{M}(u)} \right ) \cdot \\
&&\cdot \exp \left[ \int \frac{du}{2\pi} \textrm{Li}_2 \left ( -e^{-\tau E_{g}(u)+i\sigma p_{g}(u)+i\phi}
e^{X^{g}(u)} \right ) +  \int \frac{du}{2\pi} \textrm{Li}_2 \left ( -e^{-\tau E_{g}(u)+i\sigma p_{g}(u)-i\phi}
e^{X^{g}(u)} \right ) \right] \Bigr \rangle \,. \nn
\ea

\subsection{One-loop corrections}
\label{oneloop}

The purpose of this part is to disentangle the different sources of corrections to the minimal area result, \emph{i.e.} the one-loop corrections, ascribable to fermions, thus
estimating the approximations made throughout Sections \ref{ffcont} and \ref{Mes-Ins}.
%
Previously, we showed that the strong coupling limit of the fermion contribution (\ref{WL}) is given by (\ref{WLi2}), which eventually led to one node of the TBA-like equations for the hexagon. An important intermediate step was represented by the series (\ref{SingMes}) over the effective bound states $f\bar{f}$, which we called mesons. In view of this, we can identify two sources of subleading corrections, according to their different physical meaning. The analysis is most suitably carried out by considering the logarithm of the fermionic contribution to the hexagonal Wl,
\be
\log W_f\equiv \mathcal{F}_f=\mathcal{F}_M + \mathcal{F}_{free} + O\left(\frac{1}{\sqrt{\lambda}}\right) \,.
\ee
The finite term $\mathcal{F}_{free}$ represents the contribution due to unbound fermions, while $\mathcal{F}_M$ (growing as $\sqrt{\lambda}$ at strong coupling)
\be
\mathcal{F}_M=-\sqrt{\lambda}\frac{A_M}{2\pi} + \mathcal{F}_M^{sub} + O\left(\frac{1}{\sqrt{\lambda}}\right) \,,
\ee
includes $A_M$, which stands for the contribution of mesons to the minimal area, along with the finite corrections $\mathcal{F}_M^{sub}$. To sum up, the full fermion contribution enjoys the expansion
\be
\mathcal{F}_f=-\sqrt{\lambda}\frac{A_M}{2\pi} + \mathcal{F}_{free} + \mathcal{F}_M^{sub} + O\left(\frac{1}{\sqrt{\lambda}}\right) \,,
\ee
where the subleading terms are discussed in more detail below.

\ \medskip\ \\
\textbf{$\bullet$ Unbound fermions contributions $\mathcal{F}_{free}$}\\
As mentioned in the previous discussion, the first source of corrections is represented by unbound fermions: in fact, while obtaining the series over mesons (\ref{SingMes}) from the couples $f\bar{f}$, we neglected the contribution given by the interval $\mathcal{I}$ (cf. Figure \ref{Cfigura2}) as it is subleading. In more details, the main result of Section \ref{ffcont} is (\ref{conj}), which represents the leading contribution $I_n^{closed}$ to the integral over the antifermionic rapidities (\ref{I_n}) that, once plugged into (\ref{Meson}), give the way to the bound state interpretation. The remaining part $I_n^r$, ascribable to the fact that our original contour $\mathcal{C}_S$ is not closed, has to be considered if we want to study the subleading corrections. The name $\mathcal{F}_{free}$ follows from the fact that, in $I_n^r$, the rapidities are not arranged in strings, as the interval $\mathcal{I}$ is open and we do not pick up the residues.

\ \medskip\ \\
\textbf{$\bullet$ Correction from the mesons}\\
Further corrections arise from the meson series $W_M$ (\ref{WFre}). They share, to some extent, the same form as those from the Nekrasov function, which have been computed by \cite{BouFio1,BouFio2}.
More specifically, the Fredholm determinant representation for the meson series (\ref{WFre}) is exact,
\be\label{WFre1}
W_M = \left\langle\det\left(1 + M'\right)\right\rangle = \left\langle \exp\left[\sum_{n=1}^{\infty}\frac{(-1)^{n+1}}{n}\textit{Tr}(M'^n)\right]\right\rangle \, ,
\ee
provided we replace the matrix $M$ with a corrected version $M'$, which takes into account the propagator evaluated for coinciding rapidities $u_i=u_j$
\be
M'(u_i,u_j) = \frac{\left[\hat{\mu}_M(u_i-i)e^{X(u_i)}\hat{\mu}_M(u_j-i)e^{X(u_j)}\right]^{1/2}}{u_i-u_j-i} \left[P^{(MM)}_{reg}(u_i-i|u_i-i)P^{(MM)}_{reg}(u_j-i|u_j-i)\right]^{1/2}
\ee
\emph{i.e.} absorbing the diagonal terms in the Gaussian identity. This corresponds to a change in the measure $\hat{\mu}'_M(u)=\hat{\mu}_M(u)P^{(MM)}_{reg}(u|u)$: in fact, this contribution was previously disregarded, as subleading. Formula (\ref{WFre1}) can be rewritten in a customary path integral fashion
\be
W_M=\int DX e^{-S[X]}, \quad S[X]=S_{kin}[X] - \sum_{n=1}^{\infty}\frac{(-1)^{n+1}}{n}\textit{Tr}(M'^n) \,,
\ee
where $S_{kin}$ is the usual kinetic term of the action containing the inverse propagator. There are two source of corrections to $\mathcal{F}_M\equiv \log W_M$ in the strong coupling: the first one from the subleading of the saddle point (a functional determinant), the second from the expansion of the action
\be
S[X]=\sqrt{\lambda}S^{(0)} + S^{(1)} + O\left(\frac{1}{\sqrt{\lambda}}\right) \,.
\ee
In the end, we obtain the expansion
\be
\mathcal{F}_M=-\sqrt{\lambda}S_c^{(0)} - S_c^{(1)} -\frac{1}{2}\log\det H_c^{(0)}+ O\left(\frac{1}{\sqrt{\lambda}}\right) \,,
\ee
where the subscript $c$ stands for critical, \emph{i.e.} computed imposing the equation of motion $\frac{\delta S^{(0)}}{\delta X(u)}=0$, while the matrix $H^{(0)}$ is the Hessian of the leading part of the action.

The leading action $S^{(0)}$ has been obtained previously, see formula (\ref{actionM}), through several approximations. As the main contribution comes from the region with large rapidities, we neglected the shifts of $-i$ inside the mesonic quantities $\hat{
\mu}_M$, $P_{reg}^{(MM)}$. Furthermore, we used the leading formula for the measure $\hat{
\mu}_M \simeq -1$ and parametrized $P_{reg}^{(MM)}$ in terms of the kernel $K_M(\theta,\theta')$ depending on the hyperbolic rapidity $\theta$, see (\ref{Wlong}). Another main approximation concerns the evaluation of the trace in (\ref{TrLeading}), where the integration contour has been closed eventually reproducing, upon summation, the dilogarithm function. In order to evaluate $S^{(1)}$, all these effects have to be taken into account.

\section{Scalars beyond the hexagon}
\label{scalars}
\setcounter{equation}{0}

As a well established fact \cite{AM}, at strong coupling gluons and fermions (through the formation of mesons) decouple from scalars.
In the present section the focus will turn to the latter: in fact, we aim at extending the study undertaken in \cite{BFPR2, BFPR3} for the hexagon to polygons with a larger number of sides $N>6$, eventually getting in touch with \cite{BSV4}.\\
In the strong coupling limit, scalars become relativistic particles with mass $m \sim \lambda^{1/4} e^{-\frac{\sqrt{\lambda}}{4}}$ and rapidity $\theta _i=\frac{\pi}{2}u_{h,i}$.
Their contribution $\mathcal{W}_N$ to the $N$-gonal Wilson loop can be regarded as a $(N-4)$-point function of a twist operator $\mathcal{P}$ defined on the $2d$ $O(6)$ non-linear $\sigma$-model \cite{BSV4}
\be\label{Wcorr}
\mathcal{W}_N(\tau_1,\sigma_1;\cdots;\tau_{N-5},\sigma_{N-5};m)=\langle 0|\mathcal{P}(w_1)\cdots\mathcal{P}(w_{N-4})|0\rangle \, ,
\ee
where $w_{i+1}-w_{i}=(\sigma _i,\tau _i)$, with the set of cross ratios $\{\tau_i,\sigma_i,\phi_i\}_{i=1,\dots,N-5}$ fixing the
geometry of the polygon. Then,
the insertions of $N-5$ identities inside the correlator (\ref{Wcorr}) allow $\mathcal{W}_N$ to be recast as
\ba\label{WN}
\mathcal{W}_N&=&\sum_{n_1,\cdots,n_{N-5}=0}^{\infty}\prod_{l=1}^{N-5}\frac{1}{(2n_l)!}\int\prod_{l=1}^{N-5}\prod_{i_l=1}^{2n_l}\left(\frac{d\theta_{i_l}^{(l)}}{2\pi}e^{-m\tau_l\sum_{i_l=1}^{2n_l}\cosh\theta_{i_l}^{(l)}}e^{+im\sigma_l\sum_{i_l=1}^{2n_l}\sinh\theta_{i_l}^{(l)}}\right)\cdot \\
&&\cdot G^{(2n_1,\cdots ,2n_{N-5})}(\vec{\theta}^{(1)};\vec{\theta}^{(2)};\cdots ;\vec{\theta}^{(N-6)};\vec{\theta}^{(N-5)}) \, ,  \nn
\ea
where we made use of the shorthand notation $\vec{\theta}^{(l)}=(\theta^{(l)}_1,\cdots,\theta^{(l)}_{2n_l})$ for the rapidities of the $l$-th pentagon and introduced the functions $G^{(2n_1,\cdots ,2n_{N-5})}$ through their dependence on the matrix elements of $\mathcal{P}$:
\ba\label{multif}
G^{(2n_1,\cdots ,2n_{N-5})}&=&\sum_{j^{(1)}_1,\cdots ,j^{(1)}_{2n_1}}\cdots\sum_{j^{(N-5)}_1,\cdots ,j^{(N-5)}_{2n_l}}\langle 0|\mathcal{P}|\phi_{j^{(1)}_1}(\theta^{(1)}_1)\cdots \phi_{j^{(1)}_{2n_1}}(\theta^{(1)}_{2n_1})\rangle\cdots \nn \\
&& \langle \phi_{j^{(N-5)}_1}(\theta^{(N-5)}_1)
\cdots \phi_{j^{(N-5)}_{2n_{N-5}}}(\theta^{(N-5)}_{2n_{N-5}})|\mathcal{P}|0\rangle \, ,
\ea
with $|\phi_{j^{(l)}_1}(\theta^{(l)}_1)\cdots \phi_{j^{(l)}_{2n_1}}(\theta^{(l)}_{2n_1})\rangle$ a complete set of states for the $l$-th
pentagon. Choosing the normalisation $\langle 0|\mathcal{P}|0\rangle=1$, from (\ref{multif}) we can infer that, whenever one or more intermediate states correspond to the vacuum, $G^{(2n_1,\cdots ,2n_{N-5})}$ splits in terms of $G$ functions relative to polygons with lower number of sides, $i.e.$ it can be rewritten as a product of $G$'s with fewer superscripts. To give a few concrete examples, one can observe that under those assumptions, the function $G^{(2n,2m)}$ for the heptagon ($N=7$) coincides with a $G^{(2n)}$, already defined for the hexagon:
\be
G^{(2n,0)}(\theta_1,\cdots,\theta_{2n};\emptyset)=G^{(0,2n)}(\emptyset;\theta_1,\cdots,\theta_{2n})=G^{(2n)}(\theta_1,\cdots,\theta_{2n}) \,;
\ee
similarly, some functions $G^{(2n,2m,2l)}$, making their appearance in $\mathcal W_8$, can be expressed in terms of functions already present in the expansion (\ref{Wcorr}) for hexagons and heptagons, namely:
\ba
&& G^{(2n,0,0)}=G^{(0,2n,0)}=G^{(0,0,2n)}=G^{(2n)} \nn\\
&& G^{(2n,2m,0)}=G^{(0,2n,2m)}=G^{(2n,2m)} \\
&& G^{(2n,0,2m)}=G^{(2n)}G^{(2m)} \nn
\ea
(rapidities omitted for the sake of compactness).
A big deal of information can be obtained considering the logarithm of (\ref{Wcorr})
\small
\ba\label{FN}
\mathcal{F}_N&\equiv & \log\mathcal{W}_N=\sum^{\infty}_{\stackrel{(n_1,\cdots,n_{N-5})\neq (0,\cdots,0)}{n_1,\cdots,n_{N-5}=0}}\prod_{l=1}^{N-5}\frac{1}{(2n_l)!}\int\prod_{l=1}^{N-5}\prod_{i_l=1}^{2n_l} \left(\frac{d\theta_{i_l}^{(l)}}{2\pi}e^{-m\tau_l\displaystyle\sum_{i_l=1}^{2n_l}\cosh\theta_{i_l}^{(l)}}e^{+im\sigma_l\displaystyle\sum_{i_l=1}^{2n_l}\sinh\theta_{i_l}^{(l)}}\right)\cdot\nn \\
&&\cdot g^{(2n_1,\cdots ,2n_{N-5})}(\vec{\theta}^{(1)};\vec{\theta}^{(2)};\cdots ;\vec{\theta}^{(N-6)};\vec{\theta}^{(N-5)}) \equiv \sum^{\infty}_{\stackrel{(n_1,\cdots,n_{N-5})\neq (0,\cdots,0)}{n_1,\cdots,n_{N-5}=0}} \mathcal{F}_N^{(2n_1,....,2n_{N-5})}  \,.
\ea
\normalsize
The main advantage resides in the features of the functions $g^{(2n_1,\cdots ,2n_{N-5})}$, as `connected' counterparts of the $G$'s (the relation between the two sets of functions is discussed in Appendix \ref{app-conn}). Aside from a mild asymptotic behaviour\footnote{We refer the reader to \cite{BFPR2,BFPR3} for a throughout analysis of the hexagon case $\mathcal W_6$.}, the connected functions benefit from dramatic simplifications when some of the particle indices are zero, $i.e.$ some of the intermediate state in (\ref{multif}) correspond to the vacuum. Indeed, one finds
\ba
&& g^{(2n_1,\cdots ,2n_{k},0,\cdots,0)}=g^{(2n_1,\cdots ,2n_{k})} \qquad g^{(0,\cdots,0,2n_k,\cdots ,2n_{N-5})}=g^{(2n_k,\cdots ,2n_{N-5})} \label{decg7}\\
&& g^{(\cdots , 2n, 0,0,\cdots ,0,0,2m,\cdots)}=0 ,  \quad m,n\neq 0 \,. \label{zero_int}
\ea
In particular, (\ref{zero_int}) means that, whenever one of the internal indices is null, the connected function vanishes: this property can be easily understood recalling the widely known fact from statistical field theory that only connected graphs contribute to the logarithm of the partition function.

\subsection{A recursion formula for polygons}

We are now able to establish a recursion formula that elucidates how a $N$-gonal Wilson loop $\mathcal{W}_N$ can be related to polygons with fewer number of sides: this formula grows effective as the number of sides increases.\\
We first approach the heptagon, and split the series for the logarithm into three contributions
\be
\mathcal{F}_7=\sum_{(n,m)\neq (0,0)}^{\infty}\mathcal{F}_7^{(2n,2m)}=\sum_{n=1}^{\infty}\mathcal{F}_7^{(2n,0)}(\tau_1,\sigma_1) + \sum_{n=1}^{\infty}\mathcal{F}_7^{(0,2n)}(\tau_2,\sigma_2) + \sum_{n,m=1}^{\infty}\mathcal{F}_7^{(2n,2m)}(\tau_1,\sigma_1;\tau_2,\sigma_2) \,.
\ee
The property (\ref{decg7}) entails that
\be
\mathcal{F}_7^{(2n,0)}(\tau_1,\sigma_1;\tau_2,\sigma_2)=\mathcal{F}_6^{(2n)}(\tau_1,\sigma_1), \quad \mathcal{F}_7^{(0,2n)}(\tau_1,\sigma_1;\tau_2,\sigma_2)=\mathcal{F}_6^{(2n)}(\tau_2,\sigma_2) \,,
\ee
hence the heptagon turns out as the sum of two hexagons plus infinitely many `genuinely heptagonal' corrections
\be
\mathcal{F}_7(\tau_1,\sigma_1;\tau_2,\sigma_2)=\mathcal{F}_6(\tau_1,\sigma_1) + \mathcal{F}_6(\tau_2,\sigma_2) + \sum_{n,m=1}^{\infty}\mathcal{F}_7^{(2n,2m)}(\tau_1,\sigma_1;\tau_2,\sigma_2) \,.
\ee
Growing in complexity and turning to the octagon, we find out that the property (\ref{zero_int}) is responsible for the terms
$\mathcal{F}_8^{(2n,0,2m)}$ to disappear from the series (\ref{FN}), so that we obtain
\be\label{oct}
\mathcal{F}_8 = \sum_{n=1}^{\infty}\left(\mathcal{F}_8^{(2n,0,0)} + \mathcal{F}_8^{(0,2n,0)} + \mathcal{F}_8^{(0,0,2n)}\right) + \sum_{n,m=1}^{\infty}\left(\mathcal{F}_8^{(2n,2m,0)}+\mathcal{F}_8^{(0,2n,2m)}\right)  +\sum_{n,m,l=1}^{\infty}\mathcal{F}_8^{(2n,2m,2l)} \,.
\ee
It is thus apparent how the octagon can computed as the superposition of two heptagons, where the overlapping hexagon shall be subtracted in order to avoid its double-counting:
\be
\mathcal{F}_8 =\mathcal{F}_7(\tau_1,\sigma_1;\tau_2,\sigma_2) + \mathcal{F}_7(\tau_2,\sigma_2;\tau_3,\sigma_3)-\mathcal{F}_6(\tau_2,\sigma_2) + \sum_{n,m,l=1}^{\infty}\mathcal{F}_8^{(2n,2m,2l)} \,;
\ee
the latter term is an infinite collections of `purely octagonal' terms, in that they are absent in $\mathcal F_7$ and $\mathcal F_6$.\\
These formul\ae\, find a straightforward generalisation to arbitrary $N$ into the remarkable recursion relation
\ba\label{RecF-ex}
&&\mathcal{F}_{N}(\tau_1,\sigma_1;\dots;\tau_{N-5},\sigma_{N-5}) = \mathcal{F}_{N-1}(\tau_1,\sigma_1;\dots;\tau_{N-6},\sigma_{N-6})+\mathcal{F}_{N-1}(\tau_2,\sigma_2;\dots;\tau_{N-5},\sigma_{N-5})- \nn\\
&& -\mathcal{F}_{N-2}(\tau_2,\sigma_2;\dots;\tau_{N-6},\sigma_{N-6}) + \sum_{n_1,\cdots,n_{N-5}=1}^{\infty}\mathcal{F}_{N}^{(2n_1,\cdots,2n_{N-5})}(\tau_1,\sigma_1;\dots;\tau_{N-5},\sigma_{N-5}) \,,
\ea
which enjoys a suggestive interpretation: any $N$-gon results from the composition of two $(N-1)$-gons, stripped of the overlapping $(N-2)$-gon, plus infinite corrective terms.

A crucial observation concerns the behaviour of $\mathcal F_N$ in the strong coupling short distance (corresponding to $m\to 0$), in that the purely $N$-gonal terms $\mathcal{F}_N^{(2n_1,...,2n_{N-5})}$ enjoy the asymptotic expansion
\be\label{F_Nexp}
\mathcal{F}_N^{(2n_1,...,2n_{N-5})}=J_N^{(2n_1,...,2n_{N-5})}\log (1/m) + s_N^{(2n_1,...,2n_{N-5})}\log\log (1/m) + O(1) \,:
\ee
this claim will be addressed to in Appendix \ref{N-gonal_corrections}, by taking into account the simplest case, $i.e.$ heptagonal terms, then the general case as well. In the asymptotic series (\ref{F_Nexp}), the cross ratios contribute only to the finite term $O(1)$ and not to the divergent orders, so that $J$ and $s$ do not actually depend on the geometry of the loop. A recursion formula for the coefficients $J_N$ arises:
\be\label{RecJs}
J_{N} =  2J_{N-1}-J_{N-2} + \sum_{n_1,\cdots,n_{N-5}=1}^{\infty}J_{N}^{(2n_1,\cdots,2n_{N-5})}
\ee
where we have
\be
J_N=\sum^{\infty}_{\stackrel{(n_1,\cdots,n_{N-5})\neq (0,\cdots,0)}{n_1,\cdots,n_{N-5}=0}}J_N^{(2n_1,...,2n_{N-5})}, \quad s_N=\sum^{\infty}_{\stackrel{(n_1,\cdots,n_{N-5})\neq (0,\cdots,0)}{n_1,\cdots,n_{N-5}=0}}s_N^{(2n_1,...,2n_{N-5})} \,.
\ee
Finally, through the relation
\be
\log (1/m)=\frac{\sqrt{\lambda}}{4}-\frac{1}{4}\log\sqrt{\lambda}  + O(1) \,,
\ee
$\mathcal F_N$ can be expressed in terms of the coupling constant $\sqrt{\lambda}$ as
\ba\label{FNpar}
\mathcal{F}_N &=& J_N\log(1/m) + s_N\log\log(1/m) + O(1) =\\
&=&\frac{J_N}{4}\sqrt{\lambda} + \left(s_N-\frac{J_N}{4}\right)\log\sqrt{\lambda} + O(1) \,.\nn
\ea

\subsection{Solution to the recursion formula}

Some simple considerations on the recursion formula (\ref{RecJs}) can yield a good deal of information about the leading order of the logarithm of $\mathcal W_N$. Upon gathering all the purely $N$-gonal contributions into the collective quantity $\delta_N$ as
\be
\delta_N\equiv \sum_{n_1,\cdots,n_{N-5}=1}^{\infty}J_{N}^{(2n_1,\cdots,2n_{N-5})} \,,
\ee
the relation (\ref{RecJs}) assumes the form
\be\label{Recdelta}
J_{N} = 2J_{N-1}-J_{N-2} + \delta_N \,,
\ee
which is compatible with the solution
\be\label{Fdelta}
J_N = \sum_{n=6}^{N}(N+1-n)\delta_n \,,
\ee
inferred by iteratively solving (\ref{Recdelta}), once the initial conditions $J_4=J_5=0$ are imposed (as square and the pentagon are trivial
). Some precious insights about the form of $J_N$ emerge in large $N$ limit, provided we assume that $\delta_n$ decreases with $n$ fast enough for the series (\ref{Fdelta}) to make sense (we shall verify \textit{a posteriori} the consistency of this condition). In fact, under these assumptions (\ref{Recdelta}) simplifies to $J_N -2J_{N-1}+J_{N-2}=0$, which in turn shall be regarded as a discrete realisation of $\partial^2_NJ_N=0$ and, consequently, entails a linear growth (in $N$) for the solution $J_N$,
\be\label{largeN}
J_N = aN+b +O(1/N) \,,
\ee
where the expansion (\ref{Fdelta}) suggests the coefficients $a,\,b$ to take the form:
\be\label{ab}
a=\sum_{n=6}^{\infty} \delta_n, \quad b=\sum_{n=6}^{\infty} (1-n)\delta_n \,.
\ee
With an educated guess, we suppose the simplest form for the $O(N^{-1})$ corrections in (\ref{largeN}), namely
\be
J_N = aN+b + \frac{c}{N} \,,
\ee
so that the recursion relation (\ref{Recdelta}) allows us to fix $\delta_N$ up to a constant $c$
\be\label{deltaN}
\delta_N=\frac{2c}{N(N-1)(N-2)} \,:
\ee
it is worth to point out that (\ref{deltaN}) exhibits a cubic decay, ensuring the convergence of the series $a$ and $b$ (\ref{ab}). Furthermore, the latter series can be re-summed into
\be
b=-\frac{9}{20}c, \quad a=\frac{1}{20}c \,,
\ee
eventually leading (\ref{Fdelta}) to
\be\label{JNc}
J_N=\frac{c}{20}\frac{(N-4)(N-5)}{N} \,,
\ee
which reproduces the analogous result by \cite{BSV4},
up to a prefactor that we can set to $c=\frac{5}{3}$ by requiring $J_6=1/36$ \cite{BSV4}.\\
An alternative, simpler way to (\ref{JNc}) arises from demanding the linear progression (\ref{largeN}) for large $N$, then imposing that the only zeroes of $J_N$ (as a function of $N$) occur for $J_4=J_5=0$. The simplest rational solution is then
\be
J_N=\alpha\frac{(N-4)(N-5)}{N} \,,
\ee
where the constant $\alpha$ gets determined by the hexagon again, $i.e.\ \alpha=3J_6$. The argument can be straightforwardly extended to $s_N$ as well:
\be
J_N=3\frac{(N-4)(N-5)}{N}J_6, \quad s_N=3\frac{(N-4)(N-5)}{N}s_6 \,.
\ee

\section{Conclusions and perspectives}
\label{conc}
\setcounter{equation}{0}

The strong coupling limit $\lambda\to\infty$ of the null polygonal Wilson loops/scattering amplitudes in $\mathcal{N}=4$ SYM consists of two different contributions of the same order. One is due to the non-perturbative string dynamics on the sphere $S^5$ and, on the gauge theory side, is reproduced by the scalars appearing in the OPE series. This surprising effect was proposed in \cite{BSV4} and extensively analysed in \cite{BFPR2, BFPR3} for the hexagon. The other, and better understood, is the classical string contribution obtained by a minimal area problem whose solution yields a set of TBA equations \cite{Anope,TBuA,YSA, Hatsuda:2010cc}. On the gauge theory side, it comes from fermions and gluons and it has been reproduced, by a re-summation of the OPE series, in \cite{FPR} for the hexagon and extended to any polygon in \cite{BFPR1}. A crucial assumption was the fact that fermions contribute only through effective bound states fermion-antifermion, singlet under the residual $SU(4)$ R-symmetry. In this paper we definitely prove this hypothesis, by extending the proof of
\cite{BFPR1} to any number $n$ of couples fermion-antifermion.
As a first step towards this goal, in Section \ref {fermi} we dealt with the involved $SU(4)$ matrix structure, which is the main obstacle to the re-summation of the series. We computed, adapting the method of the Young tableaux previously developed for the scalars in \cite{BFPR3}, the multiple integrals defining the matrix part. The polar structure of the matrix part is thus demonstrated in general, the remaining information being encoded in some polynomials $P^{(n)}$, whose main properties are discussed in Appendix \ref {appA}. These properties are of particular importance when we compute - in Section \ref {ffcont} - the integrals over the antifermion rapidities $v_i$ by residues at the leading order in the strong coupling limit, since it turned out that the remaining fermion rapidities $u_i$ are those of effective bound states $f\bar{f}$, whose pentagonal transitions are given in terms of the known fermionic ones $P^{(ff)}$, $P^{(f\bar f)}$. The series $W_M$ over these bound states (mesons) resembles that of the Nekrasov instanton partition function $\mathcal{Z}$ for $\mathcal{N}=2$ SYM, where the role of the meson rapidities is played by the instanton positions. Starting from this similarity, in Section \ref{Mes-Ins} we could find in an elegant way the strong coupling limit of $W_M$. For that limit the emergence of bound states between mesons is crucial and in that respect a major role is played by the short-range interaction which is responsible for them and, as a consequence, for the appearance of the typical dilogarithm function. This procedure finally proves the validity of the re-summation in \cite{BFPR1, FPR}. In addition, the techniques of Section \ref{Mes-Ins} may be useful to push forward the analysis of both $W_M$ and $\mathcal{Z}$ beyond the leading order: for the latter, see for instance \cite{BouFio1, BouFio2}. The full one-loop fermion contribution, however, is more complicated, as it contains also a part due to unbound fermions, which has no analogue in the $\mathcal{N}=2$ case. Some issues on that are discussed in Subsection \ref{oneloop}. As a completion of the strong coupling analysis, in Section \ref{scalars} we discussed the non-perturbative contribution from scalars. We dealt with the general polygon by means of the same set of tools used for the hexagon in \cite{BFPR2,BFPR3}, \emph{i.e.} the expansion over the connected functions. This enables us to prove the non-perturbative enhancement of the minimal area by extracting a $\sqrt{\lambda}$ factor in front of any term in the series of the logarithm. In addition, an inspiring recursion formula among polygons is found: its physical interpretation is clear and, under reasonable assumptions, the solution reproduces the same result as \cite{BSV4} for the coefficient of the leading order.

\medskip
{\bf Acknowledgements}
We thank M. de Leeuw for discussions. This project was partially supported by the grants: GAST (INFN), UniTo-SanPaolo Nr TO-Call3-2012-0088, the ESF Network HoloGrav (09-RNP-092 (PESC)), Grant-in-Aid for JSPS Fellows 16F16735, the MPNS--COST Action MP1210 and the EC Network Gatis. DF thanks the Galileo Galilei Institute for Theoretical Physics (GGI) for the hospitality and INFN for partial support during the completion of this work. 
SP is an Overseas researcher under Postdoctoral Fellowship of Japan Society for the Promotion of Science (JSPS).

\appendix

\section{On the matrix factor for the hexagon}\label{appB}
\setcounter{equation}{0}

In this appendix we find the constraints between the numbers of flux tube excitations which enter a state with a definite $SU(4)$ charge. These numbers find a remarkable explanation in terms of the Bethe equations with $SU(4)$ symmetry for the isotopic roots $a_{k}, b_{k}, c_{k}$:
\ba
1&=&\prod _{j=1}^{N_f}\frac{a_{k}-u_{j}-i/2}{a_{k}-u_{j}+i/2} \prod _{j\not=k}^{K_a}\frac{a_{k}-a_{j}+i}{a_{k}-a_{j}-i} \prod _{j=1}^{K_b}\frac{a_{k}-b_{j}-i/2}{a_{k}-b_{j}+i/2} \nn \\
1&=&\prod _{i=1}^H \frac{b_{k}-u_{h,i}-i/2}{b_{k}-u_{h,i}+i/2} \prod _{j=1}^{K_a}\frac{b_{k}-a_{j}-i/2}{b_{k}-a_{j}+i/2} \prod _{j=1}^{K_c}\frac{b_{k}-c_{j}-i/2}{b_{k}-c_{j}+i/2} \prod _{j\not=k}^{K_b}\frac{b_{k}-b_{j}+i}{b_{k}-b_{j}-i}  \nn \\
1&=&\prod _{j=1}^{N_{\bar f}}\frac{c_{k}-v_{j}-i/2}{c_{k}-v_{j}+i/2} \prod _{j\not=k}^{K_c}\frac{c_{k}-c_{j}+i}{c_{k}-c_{j}-i} \prod _{j=1}^{K_b}\frac{c_{k}-b_{j}-i/2}{c_{k}-b_{j}+i/2} \nn
\ea
where $u_i$ are the rapidities of the fermion, $v_i$ of the antifermions and $u_{h,i}$ of the scalars.
With the help of the functions
\be
\phi (x,\xi)=i \ln \frac{i\xi +x}{i\xi -x} \, , \quad \frac{d}{dx} \phi (x,\xi)=\frac{2\xi}{x^2+\xi ^2} \, ,
\ee
we introduce the counting functions
\ba
Z^{(a)}(u)&=&-\sum _{j=1}^{N_f}\phi (u-u_{j}, 1/2 ) + \sum _{j=1}^{K_a}\phi (u-a_{j}, 1)-
\sum _{j=1}^{K_b}\phi (u-b_{j}, 1/2 ) \nn \\
Z^{(b)}(u)&=&-\sum _{i=1}^H \phi (u-u_{h,i},1/2)-\sum _{j=1}^{K_a}\phi (u-a_{j}, 1/2 ) - \sum _{j=1}^{K_c}\phi (u-c_{j}, 1/2)+\sum _{j=1}^{K_b}\phi (u-b_{j}, 1 ) \nn \\
Z^{(c)}(u)&=&-\sum _{j=1}^{N_{\bar f}}\phi (u-v_{j}, 1/2 ) + \sum _{j=1}^{K_c}\phi (u-c_{j}, 1)-
\sum _{j=1}^{K_b}\phi (u-b_{j}, 1/2 )
\ea
and the related root densities
\ba
\rho ^{(a)}(u)&=&-\frac{1}{2\pi}\frac{dZ^{(a)}}{du}=-\frac{1}{2\pi}\left [ \sum _{j=1}^{K_a}\frac{2}{(u-a_{j})^2+1}-\sum _{j=1}^{K_b}\frac{1}{(u-b_{j})^2+1/4}-
\sum _{j=1}^{N_f}\frac{1}{(u-u_{j})^2+1/4} \right ] \nn \\
\rho ^{(b)}(u)&=&-\frac{1}{2\pi}\frac{dZ^{(b)}}{du}=-\frac{1}{2\pi}\left [ \sum _{j=1}^{K_b}\frac{2}{(u-b_{j})^2+1}-\sum _{i=1}^H \frac{1}{(u-u_{h,i})^2+1/4}-\sum _{j=1}^{K_a}\frac{1}{(u-a_{j})^2+1/4}- \right. \nn \\
&-& \left.  \sum _{j=1}^{K_c}\frac{1}{(u-c_{j})^2+1/4} \right ] \nn \\
\rho ^{(c)}(u)&=&-\frac{1}{2\pi}\frac{dZ^{(c)}}{du}=-\frac{1}{2\pi}\left [ \sum _{j=1}^{K_c}\frac{2}{(u-c_{j})^2+1}-\sum _{j=1}^{K_b}\frac{1}{(u-b_{j})^2+1/4}-
\sum _{j=1}^{N_{\bar f}}\frac{1}{(u-v_{j})^2+1/4} \right ]\nn
\ea
The integrations of the density over the rapidities give
\ba
&& \int _{-\infty}^{+\infty} du \rho ^{(a)}(u)=-K_a+K_b+N_f \nn \\
&& \int _{-\infty}^{+\infty} du \rho ^{(b)}(u)=-K_b+H+K_a+K_c \label {iso-rho} \\
&& \int _{-\infty}^{+\infty} du \rho ^{(c)}(u)=-K_c+K_b+N_{\bar f} \, . \nn
\ea
Now, the $SU(4)$ singlet state (zero $R$-charge) is the state defined by distributions of the isotopic roots with no holes, which means
\be
\int _{-\infty}^{+\infty} du \rho ^{(a)}(u)=K_a \, , \quad \int _{-\infty}^{+\infty} du \rho ^{(b)}(u)=K_b  \, , \quad \int _{-\infty}^{+\infty} du \rho ^{(c)}(u)=K_c \, .
\ee
If the distributions of roots contain one hole, one gets $SU(4)$ states with non zero $R$-charge. If the hole is in the distribution of $a$-roots ($c$-roots)($b$-roots), which are connected to fermions (antifermions)(scalars), the state belongs to the $SU(4)$ representation $\mathbf{4}$ ($\mathbf{\bar{4}}$)($\mathbf{6}$). Summarizing, the conditions
\be
\int _{-\infty}^{+\infty} du \rho ^{(a)}(u)=K_a + \delta_{r,1} \, , \quad \int _{-\infty}^{+\infty} du \rho ^{(b)}(u)=K_b + \delta_{r,2} \, , \quad \int _{-\infty}^{+\infty} du \rho ^{(c)}(u)=K_c + \delta_{r,3}
\ee
define states with R-charge equal to $r=0,1,2,3$, belonging to the $SU(4)$ representation $\mathbf{1},\mathbf{4},\mathbf{6},\mathbf{\bar{4}}$, respectively.
For the number of physical and isotopic excitations we get the constraints
\be
N_f=2K_a-K_b + \delta_{r,1} \, , \quad H=2K_b-K_a-K_c + \delta_{r,2} \, , \quad N_{\bar f}=2K_c-K_b + \delta_{r,3} \, , \label {n-k}
\ee
which reproduce (and prove) formul{\ae} given in \cite{BSV6}.\\
Finally, if we restrict to singlet states made up of fermions and antifermions only, from (\ref {n-k}) we find the condition
\be
N_{f}=N_{\bar f}+4(K_b-K_c) \, . \label {ferm-singl}
\ee
Then, the particular singlet states with the same number of fermions and antifermions,
considered in Sections \ref {fermi}, \ref {ffcont}, \ref {Mes-Ins} of this paper, have number of isotopic roots $K_a=K_b=K_c=N_f$, as it is easily obtained by combining (\ref {n-k}) and (\ref {ferm-singl}).

\section{Paraphernalia}
\label{appA}
\setcounter{equation}{0}

In this appendix we collect the main properties of the polynomials $\delta_{2n}$ and $P^{(n)}$, which held a decisive role to obtain some results discussed in the main text.

\subsection{Properties of the polynomials $\delta_{2n}$}
\label{Appdelta}

The representation of the $\delta_{2n}$ as a Pfaffian (\ref{deltaPf}) 
allows for the disclosure of some helpful features of these polynomials. The most apparent property is the invariance under the exchange of arguments:
\be
\delta_{2n}(x_1,\ldots,x_i,x_{i+1},\ldots,x_{2n})=\delta_{2n}(x_1,\ldots,x_{i+1},x_i,\ldots,x_{2n}) \,.
\ee
As a less trivial feature, $\delta_{2n}$ vanishes when three or more variables lie aligned on the complex plane, with the same real part and spaced by $i$\,:
\be\label{delta_align}
\delta_{2n}(x_1,x_1+i,x_1+2i,x_4,\ldots,x_{2n})=0 \,.
\ee
When two variables differ by $i$, a recursion relation can be found:
\ba\label{delta201.1}
\delta_{2n}(2,0,1,\ldots, 1)&\equiv &\delta_{2n}(x_1, x_1+i, x_3,\ldots , x_{2n})= \nn\\
&=& 2(n-1)\displaystyle\prod_{j=3}^{2n}(x_1-x_j-i)(x_1-x_j+2i)\delta_{2n-2}(x_3,\ldots , x_{2n}) \ .
\ea
The relation (\ref{delta201.1}) above can also be iterated to find a more general expression
\ba
&&\delta_{2n}(2,0,..,2,0_{2k+2},1,\ldots ,1)\equiv\delta_{2n}(x_1, x_1+i,x_3, x_3+i,.., x_{2k+1}, x_{2k+1}+i, x_{2k+3},\ldots , x_{2n})= \nn\\
&& =2^{k+1}\frac{(n-1)!}{(n-2-k)!} \displaystyle\prod_{i<j=0}^{k}[(x_{2i+1}-x_{2j+1})^2+1][(x_{2i+1}-x_{2j+1})^2+4]\cdot \nn\\ &&\cdot\displaystyle\prod_{j=2k+3}^{2n}\displaystyle\prod_{l=0}^{k}(x_{1+2l}-x_j-i)(x_{1+2l}-x_j+2i)\delta_{2n-2-2k}(x_{2k+3}, \ldots , x_{2n})
\ea
for $0\leq k \leq n-2$; at the end of the iteration, one obtains for $\delta_{2n}(x_1, x_1+i, x_2, x_2 +i, \ldots , x_{n}, x_{n}+i)$:
\ba\label{delta20}
\delta_{2n}(2,2,\ldots ,0,0)&\equiv &\delta_{2n}(x_1, x_1+i, x_2, x_2 +i, \ldots , x_{n}, x_{n}+i)= \nn\\
&=& 2^{n-1}(n-1)!\displaystyle\prod_{i<j}^{n}[(x_{i}-x_{j})^2+1][(x_{i}-x_{j})^2+4] \ .
\ea

\subsection{Properties of the polynomials $P^{(n)}$ for fermions}

A recursion relation for the polynomials on a specific configuration can be devised from (\ref{ResPimat}):
\ba\label{RecConj}
&& P^{(n)}(u_1,\cdots , u_n,u_1-2i,v_2, \cdots , v_n)=4P^{(n-1)}(u_2,\cdots , u_n,v_2, \cdots , v_n)\cdot \\
&& \cdot \displaystyle\prod_{j=2}^n (u_{1j}+i)(u_{1j}-4i)(u_1-v_j+2i)(u_1-v_j-3i) \,.\nn
\ea
Formula (\ref{RecConj}) can be iterated $k$ times ($k\leq n$) to give
\ba\label{Reck}
&& P^{(n)}(u_1,\cdots , u_n,u_1-2i,,\cdots ,u_k -2i, v_{k+1}, \cdots , v_n)=4^k P^{(n-k)}(u_{k+1},\cdots , u_n,v_{k+1}, \cdots , v_n)\cdot \nn \\
&& \cdot \displaystyle\prod_{i=1}^k\displaystyle\prod_{j=k+1}^n (u_{ij}+i)(u_{ij}-4i)(u_i-v_j+2i)(u_i-v_j-3i)\displaystyle\prod_{i<j}^k(u_{ij}^2+1)(u_{ij}^2+16) \ ;
\ea
a remarkably simple expression is obtained for $k=n$
\be\label{fundP}
P^{(n)}(u_1,\cdots , u_n,u_1-2i, \cdots , u_n-2i)=4^n\displaystyle\prod_{i<j}^n(u^2_{ij}+1)(u^2_{ij}+16) \,.
\ee
As by-products to the recursion relation (\ref{RecConj}), one finds that, under some special configurations of the arguments, the polynomials vanish:
\ba
&& P^{(n)}(u_1,\cdots , u_n,u_1-2i,u_1-3i,v_3,\cdots ,v_n)=0 \nn\\
&& P^{(n)}(u_1,\cdots , u_n,u_1-2i,u_1 +2i,v_3,\cdots ,v_n)=0 \nn\\
&& P^{(n)}(u_1,u_1+i,u_3,\cdots , u_n,u_1-2i,v_2,\cdots ,v_n)=0 \nn\\
&& P^{(n)}(u_1,u_1-4i,u_3,\cdots , u_n,u_1-2i,v_2,\cdots ,v_n)=0 \ .
\ea
To provide a concrete example, we consider $P^{(2)}$, whose explicit form is known (\ref{P1P2}): when computed on $u_1=v_1+2i$, we get
\be
P^{(2)}(u_1,u_2,u_1-2i,v_2)=4P^{(1)}(u_2,v_2)(u_{12}+i)(u_{12}-4i)(u_1-v_2+2i)(u_1-v_2-3i) \ ;
\ee
after a further substitution $v_2=u_2+2i$, the result follows
\be
P^{(2)}(u_1,u_2,u_1-2i,u_2-2i)=16(u^2_{12}+1)(u^2_{12}+16) \ .
\ee

\section{Path integral, Fredholm determinant and the Nekrasov function}
\label{NekApp}

The method outlined in Section \ref{Mes-Ins} allows us to represent the Nekrasov function $\mathcal{Z}$ as a quantum average of a Fredholm determinant. This procedure can potentially extend beyond the leading order in the NS limit, to the subleading corrections and possibly  to the finite $\epsilon$ behaviour. Retracing the same steps as Section \ref{Mes-Ins}, we first deal with a simplified case, considering only the short-range interaction. Then, the long-range potential will be treated via a Hubbard-Stratonovich transformation. Finally, the two potentials will combined into the full partition function: we will find a representation for $\mathcal{Z}$ as a sum over bound states of instantons \cite{Bou,MenYang}, and a TBA equation in the NS limit.

\subsection{Short-range interaction}

The partition function, when only short-range interactions are taken into account, reads:
\be\label{Zshort}
\mathcal{Z}_s=\sum_{n=0}^{\infty}\frac{q^n}{n!\epsilon ^n}\int\displaystyle\prod_{i=1}^n\frac{du_i}{2\pi i}Q(u_i)\displaystyle\prod_{i<j}^n\frac{u_{ij}^2}{u_{ij}^2-\epsilon^2} \,.
\ee
The integrals are closed in the upper half plane, they are properly defined as the parameter $\epsilon$ has a positive imaginary part. We now employ two alternative approaches (Mayer expansion and Fredholm determinant) to evaluate (\ref{Zshort}) and eventually obtain the leading order for $\epsilon\to 0$.

\vspace{0.5cm}

\noindent\textbf{$\bullet $\ Mayer expansion}\\
Each two-body interaction $\frac{u_{ij}^2}{u_{ij}^2-\epsilon^2} = 1+ f(u_i-u_j)$ is diagrammatically represented as a link, corresponding to $f(x)$, connecting two nodes (associated to the particles with rapidities $u_i$ and $u_j$) of a cluster. The product in (\ref{Zshort}) is then expanded into a sum over all the different $n$-node clusters $C_n$
\be
\displaystyle\prod_{i<j}^n\frac{u_{ij}^2}{u_{ij}^2-\epsilon^2} = \sum_{C_n}\displaystyle\prod_{(i,j)\in C_n}\frac{\epsilon^2}{u_{ij}^2-\epsilon^2} \,,
\ee
where $(i,j)$ stands for the link between the nodes $i$, $j$ of the cluster. Armed with these pictorial rules, we can represent the logarithm of the grand canonical partition function as a sum restricted to the connected clusters $C^c_n$, without affecting the overall form of (\ref{Zshort})
\be
F_s\equiv \ln \mathcal{Z}_s=\sum_{n=1}^{\infty}\frac{q^n}{n!\epsilon ^n}\int\displaystyle\prod_{i=1}^n\frac{du_i}{2\pi i}Q(u_i)\sum_{C^c_n}\displaystyle\prod_{(i,j)\in C^c_n}\frac{\epsilon^2}{u_{ij}^2-\epsilon^2} \,.
\ee
A cluster $C_n^c$ is said connected if every node is connected to any other through, at least, one path of links. A tree cluster $T_n$ is a connected cluster containing exactly the minimal number of links, namely $n-1$. The crucial remark is that all the connected clusters contribute to the same order, see the discussion in \cite{Bou,MenYang}. Indeed, although one would naively expect $f(u_{ij})$ to contribute at order $\epsilon^2$, more subtly for small distances a link is proportional (in a distributional sense) to a Dirac $\delta$-function, so that $f(u_{ij})\sim\epsilon\delta(u_{ij})$, $i.e.$ a lower order in $\epsilon$.\\
In the NS limit, the residues of the poles of $Q(u)$ give a subleading contribution\footnote{Clearly, they must be taken into account for the last integration on $u_n$.}, thus we can extract a factor $Q^n(u_n)$:
\be
F_s = \sum_{n=1}^{\infty}\frac{q^n}{n!\epsilon ^n}\int \frac{du_n}{2\pi i}Q^n(u_n)\int \displaystyle\prod_{i=1}^{n-1}\frac{du_i}{2\pi i}\sum_{C^c_n}\displaystyle\prod_{(i,j)\in C^c_n}\frac{\epsilon^2}{u_{ij}^2-\epsilon^2} + O(1) \,.
\ee
We are entitled to include in the sum over clusters the disconnected diagrams too, for their contribution is vanishing, hence the short-range interactions are encoded in the multiple integral \cite{Moore}
\be\label{In}
J_n(u_n) \equiv \frac{1}{n!\epsilon^{n-1}}\int \displaystyle\prod_{i=1}^{n-1}\frac{du_i}{2\pi i}\displaystyle\prod_{i<j}^n\frac{u_{ij}^2}{u_{ij}^2-\epsilon^2} = \frac{1}{n^2} \,,
\ee
so that we get
\be\label{Fshort0}
F_s = \frac{1}{\epsilon}\sum_{n=1}^{\infty}q^n\int \frac{du}{2\pi i}Q^n(u)J_n(u) + O(1) \,.
\ee
Formula (\ref{Fshort0}) finds an interpretation in terms of bound states of instantons (tied by short-range interactions), whose component all experience the same external potential $Q(u_i)$ at leading order, as $Q(u_i+n\epsilon)\simeq Q(u_i)$. The factor (\ref{In}) does not depend on the centre of the cluster $u_n$ and assumes the role of a measure, thus shaping (\ref{Fshort0}) into a dilogarithm:
\be\label{Fshort}
F_s = \frac{1}{\epsilon}\sum_{n=1}^{\infty}\frac{q^n}{n^2}\int \frac{du}{2\pi i}Q^n(u)+ O(1) = \frac{1}{\epsilon}\int\frac{du}{2\pi i}Li_2\left[qQ(u)\right] + O(1) \,.
\ee

\vspace{0.5cm}

\noindent\textbf{$\bullet $\ Fredholm determinant}\\
The short-range partition function $\mathcal{Z}_s$ enjoys an alternative representation, which allows us to find the leading order (\ref{Fshort}) without relying on the cluster expansion. This representation, valid for any $\epsilon$, is interesting by itself and could also shed light on the $\epsilon$ corrections  to the NS limit and even analyse $\mathcal{Z}_s$ for finite $\epsilon$. The key property comes from the Cauchy formula for the short-range
\be
\frac{1}{\epsilon^n}\displaystyle\prod_{i<j}^n\frac{u_{ij}^2}{u_{ij}^2-\epsilon^2} = (-1)^n \det\left(\frac{1}{u_i-u_j-\epsilon}\right) \,,
\ee
from which we can write the whole integrand as a determinant
\be\label{Fredholm}
\mathcal{Z}_s=\sum_{n=0}^{\infty}\frac{(-q)^n}{n!}\int\displaystyle\prod_{i=1}^n\frac{du_i}{2\pi i}\det_{ij} M(u_i,u_j) \,,
\ee
where the kernel $M(u_i,u_j)$ includes the potential $Q$ and reads
\be\label{M}
M(u_i,u_j)=\frac{Q^{1/2}(u_i)Q^{1/2}(u_j)}{u_i-u_j-\epsilon} \,.
\ee
The expression (\ref{Fredholm}) is the definition of the Fredholm determinant for the integral operator $M(u_i,u_j)$
\be\label{Fred}
\mathcal{Z}_s= \det(1 - qM) \,.
\ee
This formula holds for any $\epsilon$ and regardless of the functional form of $Q(u)$, as the only property we employed is the Cauchy identity for the short-range interaction. Formula (\ref{Fred}) comes in handy when we consider the logarithm $F_s$, using the identity $\log\det=\textit{Tr}\log$ and expanding  we get
\be\label{SerTrace}
F_s=\log \mathcal{Z}_s =\log\det (1-qM)=\textit{Tr}\log(1-qM)=-\sum_{n=1}^{\infty}\frac{q^n}{n}\textit{Tr}M^n \,.
\ee
The trace of an integral operator is defined as
\be
\textit{Tr}M^n \equiv \int\displaystyle\prod_{i=1}^n\frac{du_i}{2\pi i} \displaystyle\prod_{i=1}^n M(u_i,u_{i+1}) = \int\displaystyle\prod_{i=1}^n\frac{du_i}{2\pi i}Q(u_i) \displaystyle\prod_{i=1}^n \frac{1}{u_i-u_{i+1}-\epsilon}, \quad u_{n+1}\equiv u_1
\ee
Now we employ the small $\epsilon$ limit. The main contribution to the trace is given by the residues of the polar part $\frac{1}{u_i-u_{i+1}-\epsilon}$: we perform the $n-1$ integrations to obtain
\be\label{TrLead}
\textit{Tr}M^n = -\frac{1}{n\epsilon}\int\frac{du}{2\pi i}Q^n(u) + O(1) \,,
\ee
where the shifts inside the functions $Q(u+k\epsilon)$ have been neglected, as $Q$ enjoys a smooth $\epsilon\to 0$ limit. Summing the series (\ref{SerTrace}) with the leading order (\ref{TrLead}), we reproduce the result (\ref{Fshort}) previously obtained from the cluster expansion.

To conclude our analysis, we show how the leading order (\ref{Fshort}), which contains the dilogarithm function, is equivalent to a sum over bound states of instantons. 
Let us recall the partition function $\mathcal{Z}_s$ in the small $\epsilon$ limit:
\be
\mathcal{Z}_s \simeq \exp\left[\frac{1}{\epsilon}\int\frac{du}{2\pi i}Li_2\left[qQ(u)\right]\right] \,.
\ee
We expand both the dilogarithm and the exponential to get
\be
\mathcal{Z}_s=\sum_{N=0}^{\infty}\frac{1}{N!\epsilon^N}\left[\int\frac{du}{2\pi i}Li_2[qQ(u)]\right]^N=\sum_{N=0}^{\infty}\frac{1}{N!\epsilon^N}\left[\sum_{a=1}^{\infty}\int\frac{du}{2\pi i}\frac{q^a Q^a(u)}{a^2}\right]^N \,,
\ee
from which we can write the $N$-th power as a multiple sum over $a_i$
\be
\mathcal{Z}_s=\sum_{N=0}^{\infty}\frac{1}{N!\epsilon^N}\sum_{a_1=1}^{\infty}\cdots\sum_{a_N=1}^{\infty}\int\displaystyle\prod
_{i=1}^N\frac{du_i}{2\pi i}\frac{q^{a_i}Q^{a_i}(u_i)}{a_i^2} \,.
\ee
This series is, at the leading order in the small $\epsilon$ limit, equivalent to the initial definition (\ref{Zshort}) of $\mathcal{Z}_s$. Here, $N$ represents the number of composite particles, while $a_i$ tells us how many instantons are bound inside the $i$-th particle. It is worth to remark that the typical dilogarithm function appears thanks to the particular measure of the bound states $1/a^2$, see the integral (\ref{In}). 

\subsection{Long-range interaction}

In this subsection we deal with the other simplified case, where only the long-range interaction is present. This time, the partition function reads
\be\label{Zlong}
\mathcal{Z}_L=\sum_{n=0}^{\infty}\frac{q^n}{n!\epsilon ^n}\int\displaystyle\prod_{i=1}^n\frac{du_i}{2\pi i}Q(u_i)\displaystyle\prod_{i<j}^n e^{\epsilon G(u_{ij})} \,.
\ee
The difference with respect to $\mathcal{Z}_s$ is that the two-body potential is smooth in the limit $\epsilon\to 0$ and we can push the Mayer expansion all the way through. As before, we define $e^{\epsilon G(u)}\equiv 1 + \epsilon f(u)$, thus the free energy $F_L$ is the sum over all the connected clusters
\be
F_L\equiv \ln \mathcal{Z}_L=\sum_{n=1}^{\infty}\frac{q^n}{n!\epsilon ^n}\int\displaystyle\prod_{i=1}^n\frac{du_i}{2\pi i}Q(u_i)\sum_{C^c_n}\displaystyle\prod_{(i,j)\in C^c_n}\epsilon f(u_{ij}) \,.
\ee
Since we do not have a singular behaviour for $\epsilon\to 0$, there are no subtleties and the leading order is simply given by the tree clusters, which contain $n-1$ links and cancel all the powers of $\epsilon$ but one
\be\label{tree}
F_L = \sum_{n=1}^{\infty}\frac{q^n}{n!\epsilon }\int\displaystyle\prod_{i=1}^n\frac{du_i}{2\pi i}Q(u_i)\sum_{T_n}\displaystyle\prod_{(i,j)\in T_n} f(u_{ij}) + O(1) \,.
\ee
We remind that this statement is not true for the short-range, where all the connected clusters contribute at the leading order and their effect was computed by the integral (\ref{In}).

\vspace{0.5cm}

\noindent\textbf{$\bullet $\ Path integral representation}\\
An alternative method to study $\mathcal{Z}_L$ makes use of a path integral representation \cite{MenYang, BouFio1, BouFio2}. The long-range potential admits the natural interpretation of the propagator of a quantum field $X(u)$
\be
\left\langle X(u)X(v)\right\rangle \equiv \epsilon G(u-v) \,.
\ee
The Gaussian identity, extended to the functional description, leads to the important equivalence
\be\label{Hubb}
\displaystyle\prod_{i<j}^n e^{\epsilon G(u_{ij})}=\displaystyle\prod_{i<j}^n e^{\left\langle X(u_i)X(u_j)\right\rangle} = e^{-\frac{1}{2}n\epsilon G(0)} \left\langle\displaystyle\prod_{i=1}^n e^{X(u_i)}\right\rangle
\ee
which enables us to represent the two-body interaction through an average of single particle terms. This procedure is known in literature as the Hubbard-Stratonovich transformation. We define the renormalized instanton parameter as $q'=q e^{-\frac{\epsilon}{2}G(0)}$, to account for the diagonal term in (\ref{Hubb}). The partition function is thus written as an expectation value
\be\label{ZLaverage}
\mathcal{Z}_L=\left\langle \exp\left[\frac{q'}{\epsilon}\int\frac{du}{2\pi i} Q(u)e^{X(u)}\right]\right\rangle \,,
\ee
where the average of a generic functional $F[X]$ is defined by the path integral
\be
\left\langle F[X] \right\rangle \equiv \int DX F[X] \exp\left[\frac{1}{\epsilon} S_0[X]\right] , \quad S_0[X]=- \frac{1}{2}\int \frac{dudv}{(2\pi i)^2}X(u)G^{-1}(u-v)X(v) \,.
\ee
The inverse propagator satisfies
\be
\int \frac{dv}{2\pi i} G^{-1}(u-v)G(v-w)=\delta(u-w) \,,
\ee
where the Dirac delta function is defined with the normalization
\be
\int \frac{du}{2\pi i} f(u)\delta(u-v)=f(v) \,.
\ee
The partition function (\ref{Zlong}) is thus recast as a path integral
\be
\mathcal{Z}_L=\int DX \exp\left[\frac{1}{\epsilon}S[X]\right]
\ee
with the action
\be
S[X]=S_0[X] + q'\int\frac{du}{2\pi i} Q(u)e^{X(u)} \,.
\ee
We remark that the path integral representation (\ref{ZLaverage}) for (\ref{Zlong}) is valid for any $\epsilon$. However, having extracted a factor $\epsilon^{-1}$ in front of the action $S[X]$, the limit $\epsilon\to 0$ follows immediately by the saddle point approximation
\be
F_L \simeq \frac{1}{\epsilon}S[X_c]\equiv \frac{1}{\epsilon}S_c \,.
\ee
The saddle point equation comes from $\frac{\delta S[X]}{\delta X(u)}=0$, it reads\footnote{At leading order, the instanton parameter is not corrected.}
\be
qQ(u)e^{X(u)}=\int\frac{dv}{2\pi i}G^{-1}(u-v)X(v) \,,
\ee
which can be expressed in term of the direct kernel as
\be\label{eom}
X(u)=q\int\frac{dv}{2\pi i}G(u-v)Q(v)e^{X(v)} \,.
\ee
The critical action is then
\be\label{Scr}
S_c = q\int\frac{du}{2\pi i}Q(u)\left[1-\frac{1}{2}X(u)\right]e^{X(u)} \,,
\ee
where $X(u)$ satisfies the classical equation of motion (\ref{eom}).
As a check, if we expand the solution (\ref{eom}) in powers of $q$ and the substitute in (\ref{Scr}), we get the standard expansion over the connected tree clusters (\ref{tree}) for $F_L$. This method turns out useful when dealing with the full partition function $\mathcal{Z}$ and with the series of the mesons $W_M$.

\subsection{The full partition function}

Now we are ready to tackle the whole Nekrasov partition function, which contains both types of interaction discussed in the previous subsections. We recall here the formula
\be\label{Z}
\mathcal{Z}=\sum_{n=0}^{\infty}\frac{q^n}{n!\epsilon ^n}\int\displaystyle\prod_{i=1}^n\frac{du_i}{2\pi i}Q(u_i)\displaystyle\prod_{i<j}^n e^{\epsilon G(u_{ij})}\displaystyle\prod_{i<j}^n\frac{u_{ij}^2}{u_{ij}^2-\epsilon^2} \,.
\ee
We point out that the leading order in the NS limit has already been unravelled in \cite{Bou,MenYang}, mainly by means of the Mayer expansion. Here we use a different approach, which combines both the techniques introduced before and yields the leading order behaviour much faster.
To address the problem we apply in sequence the Hubbard-Stratonovich transformation and the Fredholm formula. First, we use the fluctuating field $X(u)$ to obtain a path integral representation
\be\label{Zaverage}
\mathcal{Z}=\left\langle \mathcal{Z}_s[q\to q', Q\to Qe^X]\right\rangle
\ee
which differs from (\ref{ZLaverage}), since we still have the short-range interaction to deal with. As a matter of fact, (\ref{Zaverage}) is the expectation value of a short-range partition function $\mathcal{Z}_s$, where the potential is modified by the fluctuating field through $Q(u) \to Q(u)e^{X(u)}$. Of course, the renormalized instanton coupling appears.\\
Now we can work out the short-range part with the Fredholm technique, so that we have a fluctuating matrix $M'[X]$, related to $M$ in (\ref{M}) through
\be
M'_{ij}[X]=M_{ij}\exp\left[\frac{1}{2}X(u_i) + \frac{1}{2}X(u_j)\right] \,.
\ee
The full partition function is then the expectation value of a Fredholm determinant
\be\label{ZFre}
\mathcal{Z}=\left\langle\det(1-q'M'[X])\right\rangle
\ee
which stands correct for any $\epsilon$.
In the NS limit we obtain the same result as for $\mathcal{Z}_s$ and the dilogarithm appears
\be\label{Zpath}
\mathcal{Z}\simeq \left\langle\exp\left[\frac{1}{\epsilon}\int\frac{du}{2\pi i}Li_2[qQ(u)e^{X(u)}]\right]\right\rangle = \int DX \exp\left[\frac{1}{\epsilon}S[X]\right] \,,
\ee
with the difference that the argument contains the field $X(u)$ over which we average.The total action is
\be
S[X]=-\frac{1}{2}\int\frac{dudv}{(2\pi i)^2}X(u)G^{-1}(u-v)X(v) + \int\frac{du}{2\pi i}Li_2[qQ(u)e^{X(u)}] \,.
\ee
For small $\epsilon$ the path integral is dominated by the critical action
\be
F=\ln \mathcal{Z} \simeq \frac{1}{\epsilon}S[X_c]=\frac{1}{\epsilon}S_c
\ee
coming from the saddle point, which resembles the TBA equation
\be\label{TBAlikeZ}
X(u) + \int\frac{dv}{2\pi i}G(u-v)\ln\left[1-qQ(v)e^{X(v)}\right]=0
\ee
from which the critical action follows
\be
S_c=\frac{1}{2}\int\frac{du}{2\pi i}X(u)\ln\left[1-qQ(u)e^{X(u)}\right] + \int\frac{du}{2\pi i}Li_2\left[qQ(u)e^{X(u)}\right]
\ee
that matches the critical value of the Yang-Yang functional in \cite{Bou,MenYang}.\\
As we did for the short-range, we can get the sum over bound states by expanding the dilogarithm and the exponential inside the average in (\ref{Zpath}). The generalized Gaussian identity, neglecting the diagonal term, reads
\be
\displaystyle\prod_{i<j}^N e^{a_i a_j \left\langle X(u_i)X(u_j)\right\rangle} \simeq \left\langle\displaystyle\prod_{i=1}^N e^{a_i X(u_i)}\right\rangle
\ee
and allows us to find the alternative expression of the partition function in the $\epsilon\to 0$ limit
\be\label{Zbound}
\mathcal{Z}\simeq \sum_{N=0}^{\infty}\frac{1}{N!\epsilon^N}\sum_{a_1=1}^{\infty}\cdots\sum_{a_N=1}^{\infty}\int
\displaystyle\prod_{i=1}^N\frac{du_i}{2\pi i}\frac{q^{a_i}Q^{a_i}(u_i)}{a_i^2}\displaystyle\prod_{i<j}^N e^{\epsilon a_i a_j G(u_{ij})}
\ee
as a sum over the bound states formed by the short-range interaction, which are interacting through the long range part $\epsilon a_i a_jG(u_{ij})$. The numbers $a_i$ represent the number of elementary constituent of the bound state, whose measure is proportional to $a_i^{-2}$; the long range interaction acts between any couple of elementary constituents, so that the total effect contains the multiplicity factor $a_i a_j$. The sum (\ref{Zbound}) closely resembles the series over free and bound mesons (\ref{BoundMes}), making the analogy between $\mathcal{Z}$ and $W^{(M)}$ even more manifest.

\section{Gluons and fermions}
\label{use_f}
\setcounter{equation}{0}

We collect here some useful expressions, employed in Section \ref{glufer}, to describe how bound states of gluons couple to fermions and mesons. First, it is appropriate to introduce the Zhukovsky map and its shifted version
\be
x(u)=\frac{u}{2}\left[1+\sqrt{1-\frac{4g^2}{u^2}}\right] \qquad\qquad
x^{[\pm \kappa]}(u)=x(u\pm\frac{i\kappa}{2}) \ .
\ee
The gluon-fermions pentagonal amplitudes are related to the scattering phases \cite{Bel1410}
\ba\label{PS}
\,[P^{(gf)}_a(u|v)]^2 &=& w_{gf}^{(a)}(u,v)\frac{S^{(gf)}_a(u,v)}{S^{(gf)}_a(u^\gamma,v)} \\
\,[P^{(g\bar f)}_a(u|v)]^2 &=& w_{g\bar f}^{(a)}(u,v)\frac{S^{(g\bar f)}_a(u,v)}{S^{(g\bar f)}_a(u^{\gamma},v)} \nn\\
\,[P^{(fg)}_a(v|u)]^2 &=& w_{fg}^{(a)}(v,u)\frac{S^{(fg)}_a(v,u)}{S^{(fg)}_a(v,u^{-\gamma})} \nn\\
\,[P^{(\bar fg)}_a(v|u)]^2 &=& w_{\bar fg}^{(a)}(v,u)\frac{S^{(\bar fg)}_a(v,u)}{S^{(\bar fg)}_a(v,u^{-\gamma})} \,,\nn
\ea
where the functions $w_{gf}^{(a)},\,w_{g\bar f}^{(a)}$ can be written (introducing $x_f(v)=g^2/x(v)$) as \cite{Bel1509}:
\ba
w_{gf}^{(a)}(u,v) &=& (-1)^{a+1}(u-v+\frac{i|a|}{2})\frac{x_f(v)}{x^{[+a]}(u)x^{[-a]}(u)}
\left(1-\frac{x_f(v)}{x^{[+a]}(u)}\right)^{-1}\left(1-\frac{x_f(v)}{x^{[-a]}(u)}\right)^{-1} \nn\\
w_{g\bar f}^{(a)}(u,v) &=& [w_{gf}^{(a)}(u,v)]^{-1} \,.
\ea
The pentagonal transition $P^{(gM)}_a(u|v)$ (at any coupling) involving mesons is obtained through the fusion of fermionic quantities:
\ba
&& [P^{(gM)}_a(u|v)]^2=[P^{(gf)}_a(u|v+i)P^{(g\bar f)}_a(u|v-i)]^2=[P^{(gf)}_a(u|v+i)P^{(\bar g f)}_a(u|v-i)]^2= \nn \\
&& w_{gf}^{(a)}(u,v+i) w_{\bar gf}^{(a)}(u,v-i)\frac{S^{(gf)}_a(u,v+i)}{S^{(gf)}_a(u^{\gamma},v+i)}
\frac{S^{(\bar gf)}_a(u,v-i)}{S^{(\bar gf)}_a(u^{\gamma},v-i)} \,.
\ea
In the strong coupling regime, we can provide the expressions above in a closed form. In pursuing this aim, it is worth to recall the hyperbolic rapidities for fermions
\be
v=2g\bar v=2g\coth 2\q'
\ee
and for gluons
\be
u=2g\bar u=2g\tanh 2\q \,.
\ee
Using these variables, the mirror transformation can be implemented in a easy way \textit{at strong coupling} (at finite coupling instead, its representation requires much more effort, see \cite{BSV3})
\be
\q ^{\gamma}=\q+\frac{i\pi}{2} \, .
\ee
The scattering phases at strong coupling can be written in a compact form, at accuracy $O(1/g)$ 
(setting $a$ to be a positive integer, $a=1,2,3,\dots$)
\ba
S^{(gf)}_a(\q,\q') &=& S^{(g \bar f)}_{-a}(\q,\q')=\exp \left[ \frac{ia}{4g}\frac{\sqrt{2}\cosh(\q-\q')+1}{\tanh 2\q-\coth 2\q'}\right] \nn\\
S^{(g\bar f)}_a(\q,\q') &=& S^{(g f)}_{-a}(\q,\q')=\exp \left[\frac{ia}{4g}\frac{\sqrt{2}\cosh(\q-\q')-1}{\tanh 2\q-\coth 2\q'}\right] \nn\\
S^{(gf)}_a(\q^{\gamma},\q') &=& S^{(g\bar f)}_{-a}(\q^{\gamma},\q')=\exp \left[\frac{ia}{4g}\frac{i\sqrt{2}\sinh(\q-\q')+1}{\tanh 2\q-\coth 2\q'} \right] \nn \\
S^{(g\bar f)}_a(\q^{\gamma},\q') &=& S^{(g f)}_{-a}(\q^{\gamma},\q')=\exp \left [\frac{ia}{4g}\frac{i\sqrt{2}\sinh(\q-\q')-1}{\tanh 2\q-\coth 2\q'} \right ] \nn \\
S^{(gM)}_a(\q,\q') &=& S^{(gf)}_{a}(\q,\q') S^{(g\bar f)}_a(\q,\q') = \exp \left[ \frac{ia}{2g}\frac{\sqrt{2}\cosh(\q-\q')}{\tanh 2\q-\coth 2\q'} \right] \, \nn
\ea
while for the $w$-functions it results
\be
w_{gf}^{(a)}(u,v+i) w_{\bar gf}^{(-a)}(u,v-i)=
1+O(1/g^2) \,.
\ee
Eventually, the gluon-meson pentagon amplitudes can be found:
\ba
P^{(gM)}_a(\q|\q') &=& 1-\frac{|a|}{2g}\frac{\cosh2\q\sinh 2\q'}{\sqrt{2}\cosh(2\q-2\q')}[\sinh(\q-\q')+i\cosh(\q-\q')] \\
P^{(Mg)}_a(\q'|\q) &=& 1+\frac{|a|}{2g}\frac{\cosh2\q\sinh 2\q'}{\sqrt{2}\cosh(2\q-2\q')}[\sinh(\q'-\q)+i\cosh(\q'-\q)]  \,. \nn
\ea

\section{Scalars}
\label{scal}
\setcounter{equation}{0}

To better understand the claims of Section \ref{scalars}, we provide some additional formul\ae\ regarding the scalar contribution to the polygonal Wl: in particular, we list some properties of the connected functions $g^{(2n_1,...,2n_{N-5})}$ and analyse
$\mathcal{F}_N^{(2n_1,...,2n_{N-5})}$.

\subsection{Connected functions}
\label{app-conn}

The connected functions enjoy the general form
\be\label{conn_g}
g^{(2n_1,\cdots,2n_k)}=\sum_{l=1}^{n_1+ \cdots + n_k}(-1)^{l-1}(l-1)!\sum_{\left \{ n^{(j)}_m \right \}}\sum_{d.e.}\prod_{j=1}^{l}G^{(2n^{(j)}_1,\cdots,2n^{(j)}_{k})}
\ee
where $l$ is the number of functions $G$ appearing in the product, $\left \{ n^{(j)}_m \right \}$ represents the set of different products of $l$ functions (subject to the constraint $\displaystyle\sum_{j=1}^{l}n^{(j)}_m=n_m$) and the last sum contains all the permutations (different exchanges, $d.e.$) among the equivalent rapidities.
Formula (\ref{conn_g}) and the factorisation of the $G$'s
\be\label{dec1}
G^{(2n_1,\cdots,2n_k,0,\cdots,0,2m_1,\cdots,2m_l)}=G^{(2n_1,\cdots,2n_k)}G^{(2m_1,\cdots,2m_l)}, \quad G^{(2n_1,\cdots,2n_k,0,0,\cdots,0,0)}=G^{(2n_1,\cdots,2n_k)}
\ee
entail the main properties for the connected functions
\be\label{propg1}
g^{(2n_1,\cdots,2n_k,0,0,\cdots,0,0)}=g^{(2n_1,\cdots,2n_k)}, \quad g^{(.....,2n,0,0,.....,0,0,2m,.....)}=0 \quad (m,n\neq 0) \,.
\ee

\vspace{0.5cm}

\noindent\textbf{$\bullet $\ Heptagon}\\
In order to make the formul\ae\  for the connected functions easier to visualise, we display some explicit expressions for the heptagon. Up to six particles, the non trivial heptagonal functions are $g^{(2,2)}$, $g^{(4,2)}$ and $g^{(2,4)}$, for all the remaining ones can be reduced via (\ref{propg1}) to functions already present in the hexagon. The simplest case corresponds to
\ba
g^{(2,2)}(\theta_1,\theta_2;\theta'_1,\theta'_2)&=&G^{(2,2)}(\theta_1,\theta_2;\theta'_1,\theta'_2) -G^{(2,0)}(\theta_1,\theta_2;\emptyset)G^{(0,2)}(\emptyset;\theta'_1,\theta'_2) =\\
&=& G^{(2,2)}(\theta_1,\theta_2;\theta'_1,\theta'_2) -G^{(2)}(\theta_1,\theta_2)G^{(2)}(\theta'_1,\theta'_2) \,.\nn
\ea
The six particles function $g^{(4,2)}$, $g^{(2,4)}$ are related by symmetry and are given by
\ba
&& g^{(4,2)}(\theta_1,\theta_2,\theta_3,\theta_4;\theta'_1,\theta'_2)=G^{(4,2)}(\theta_1,\theta_2,\theta_3,\theta_4;\theta'_1,\theta'_2)- G^{(2)}(\theta'_1,\theta'_2) G^{(4)}(\theta_1,\theta_2,\theta_3,\theta_4) - \\
&-& \left(G^{(2)}(\theta_1,\theta_2)G^{(2,2)}(\theta_3,\theta_4;\theta'_1,\theta'_2) + \textit{5 terms}\right) + 2G^{(2)}(\theta'_1,\theta'_2)\left(G^{(2)}(\theta_1,\theta_2)G^{(2)}(\theta_3,\theta_4) + \textit{2 terms}\right) \nn
\ea
where the parenthesis contain all the permutations of the rapidities,
making $g^{(4,2)}$ symmetric under the exchange of any of the $\theta_i$.

\vspace{0.5cm}

\noindent\textbf{$\bullet $\ Octagon}\\
From $N=8$ the second property of (\ref{propg1}) starts to play an important role, as many contributions disappear.
Since $g^{2,0,2}=g^{4,0,2}=g^{2,0,4}=0$, the first non-zero octagonal function is
\be
g^{(2,2,2)}=G^{(2,2,2)}-G^{(2,2,0)}G^{(0,0,2)}-G^{(2,0,2)}G^{(0,2,0)}-G^{(0,2,2)}G^{(2,0,0)} + 2G^{(2,0,0)}G^{(0,2,0)}G^{(0,0,2)} \,,
\ee
no permutations involved since there are at most two rapidities in each set. More explicitly, making use of (\ref{dec1}) we find
\ba
&& g^{(2,2,2)}(\theta_1,\theta_2;\theta'_1,\theta'_2;\theta''_1,\theta''_2) = G^{(2,2,2)}(\theta_1,\theta_2;\theta'_1,\theta'_2;\theta''_1,\theta''_2) - G^{(2,2)}(\theta_1,\theta_2;\theta'_1,\theta'_2)G^{(2)}(\theta''_1,\theta''_2) - \nn \\
&& - G^{(2)}(\theta_1,\theta_2)G^{(2,2)}(\theta'_1,\theta'_2;\theta''_1,\theta''_2) + G^{(2)}(\theta_1,\theta_2)G^{(2)}(\theta'_1,\theta'_2)G^{(2)}(\theta''_1,\theta''_2)
\ea

\subsection{N-gonal corrections}\label{N-gonal_corrections}

In the following, the validity of the expansion (\ref{F_Nexp}) for $\mathcal{F}_N^{(2n_1,....,2n_{N-5})}$ will be shown, as well as a formula for the leading coefficient $J_N^{(2n_1,....,2n_{N-5})}$ will be found.

\vspace{0.5cm}

\noindent\textbf{$\bullet $\ Heptagon}\\
The argument of \cite{BFPR2, BFPR3} will be extended to the heptagon ($N=7$), choosing as a starting expression
\be
\mathcal{F}_7^{(2n,2m)}=\frac{1}{(2n)!(2m)!}\int\prod_{i=1}^{2n}\frac{d\theta_i}{2\pi}\prod_{j=1}^{2m}\frac{d\theta'_j}{2\pi}e^{-m\tau_1\sum_i\cosh\theta_i}e^{-m\tau_2\sum_j\cosh\theta'_j+im\sigma_2\sum_j\sinh\theta'_j}g^{(2n,2m)}
\ee
where, for simplicity, we eliminated the cross ratio $\sigma_1$ by a rotation.
The connected function $g^{(2n,2m)}$ depends on the differences $\theta_{ij}$, $\theta'_{ij}$ and $\theta_i-\theta'_j$, thus we define $\alpha_i\equiv \theta_i-\theta_1$, $\alpha'_j\equiv\theta'_j-\theta_1$, with $i=2,\dots,2n$ and $j=1,\dots,2m$. Switching to the variables
$\theta_1\equiv\theta,\,\alpha_i$ and $\alpha'_j$, the dependence on $\q$ is stripped from $g^{(2n,2m)}$, so (with a slight abuse of notation)
we write
\ba\label{F7theta}
\mathcal{F}_7^{(2n,2m)}&=&\frac{1}{(2n)!(2m)!}\int\prod_{i=2}^{2n}\frac{d\alpha_i}{2\pi}\prod_{j=1}^{2m}\frac{d\alpha'_j}{2\pi}g^{(2n,2m)}(\alpha_2,\cdots,\alpha_{2n};\alpha'_1,\cdots,\alpha'_{2m})\cdot \\
&\cdot & \int d\theta \exp\left[-m\tau_1\xi\cosh\left(\theta+\eta \right) - m\tau_2\xi'\cosh\left(\theta+\eta' \right) + im\sigma_2\xi'\sinh\left(\theta+\eta'\right)\right]   \nn
\ea
where $\xi$, $\xi'$, $\eta$ and $\eta'$ are functions of $\alpha_i$, $\alpha'_j$:
\small
\be
 1 +\sum_{i=2}^{2n}\cosh\alpha_i=\xi\cosh\eta , \quad \sum_{i=2}^{2n}\sinh\alpha_i=\xi\sinh\eta, \quad
 \sum_{j=1}^{2m}\cosh\alpha'_j=\xi'\cosh\eta' , \quad \sum_{j=1}^{2m}\sinh\alpha'_j=\xi'\sinh\eta' \,.
\ee
\normalsize
The integral on $\theta$ in (\ref{F7theta}) is a more complicated version of the Bessel function $2 K_0(z\xi)=\int d\theta e^{-z\xi\cosh\theta}$, but for small $m$ they behave similarly and the leading $m\to 0$ (divergent) term can be extracted by trading the exponentials for a finite integration volume $-\log(1/m)<\theta<\log (1/m)$
\ba\label{mto0}
&&\int d\theta \exp\left[-m\tau_1\xi\cosh\theta - m\tau_2\xi'\cosh\left(\theta+\eta' -\eta \right) + im\sigma_2\xi'\sinh\left(\theta+\eta' -\eta\right)\right]\simeq \nn \\
&& \simeq \int_{-\log(1/m)}^{\log(1/m)}d\theta=2\log(1/m) \,,
\ea
to get an explicit expression for the coefficient as
\be
J_7^{(2n,2m)}=\frac{2}{(2n)!(2m)!}\int\prod_{i=2}^{2n}\frac{d\alpha_i}{2\pi}\prod_{j=1}^{2m}\frac{d\alpha'_j}{2\pi}g^{(2n,2m)}(\alpha_2,\cdots,\alpha_{2n};\alpha'_1,\cdots,\alpha'_{2m}) \,.
\ee
The subleading divergence requires the introduction of a cutoff, whose removal yields a term proportional to $\log\log(1/m)$. This has been extensively discussed in \cite{BFPR3} for the hexagon case.

\vspace{0.5cm}

\noindent\textbf{$\bullet $\ General case $N>7$}\\
The procedure just described can be generalised to polygons with an arbitrary number of sides.
Again, we can handle the expression
\ba\label{FNnnn}
\mathcal{F}_N^{(2n_1,\dots,2n_{N-5})} &=& \prod_{l=1}^{N-5}\frac{1}{(2n_l)!}\int\prod_{l=1}^{N-5}\prod_{i_l=1}^{2n_l}\left(\frac{d\theta_{i_l}^{(l)}}{2\pi}e^{\displaystyle-m\tau_l\sum_{i_l=1}^{2n_l}\cosh\theta_{i_l}^{(l)}+im\sigma_l\sum_{i_l=1}^{2n_l}\sinh\theta_{i_l}^{(l)}}\right)\cdot \nn\\
&&\cdot g^{(2n_1,\cdots ,2n_{N-5})}(\vec{\theta}^{(1)};\cdots;\vec{\theta}^{(N-5)})
\ea
making use of the variables $\alpha_i^{(l)}\equiv \theta_i^{(l)}-\theta_1^{(1)},\ \theta_1^{(1)}\equiv\theta$, and introducing
$\xi,\,\eta,\,\xi^{(l)},\,\eta^{(l)}$
via the relations
\ba
&& 1 +\sum_{i=2}^{2n_1}\cosh\alpha_i^{(1)}=\xi\cosh\eta \ , \qquad\qquad \sum_{i=2}^{2n_1}\sinh\alpha_i^{(1)}=\xi\sinh\eta \\
&& \sum_{j=1}^{2n_l}\cosh\alpha_j^{(l)}=\xi^{(l)}\cosh\eta^{(l)} \ ,\qquad\qquad \sum_{j=1}^{2n_l}\sinh\alpha^{(l)}_j=\xi^{(l)}\sinh\eta^{(l)}\,.\nn
\ea
The quantity (\ref{FNnnn}) can be thus recast into
\small
\ba
&&\mathcal{F}_N^{(2n_1,\dots,2n_{N-5})} = \prod_{l=1}^{N-5}\frac{1}{(2n_l)!}\int\frac{d\q}{2\pi}\int\prod_{l=1}^{N-5}\prod_{i_l=1}^{2n_l}
\left[\frac{d\alpha_{i_l}^{(l)}}{2\pi}\,e^{\displaystyle-m\tau_l\xi^{(l)}\cosh(\q+\eta^{(l)})+im\sigma_l\sinh(\q+\eta^{(l)})}\right]\cdot \nn\\
&&\cdot\ e^{-m\tau_1\xi\cosh\left(\theta+\eta \right)}\,
g^{(2n_1,\cdots ,2n_{N-5})}(\alpha_2^{(1)},\cdots,\alpha_{2n_1}^{(1)};\alpha_1^{(2)}\cdots\alpha_{2n_{N-5}}^{(N-5)}) \,,
\ea
\normalsize
which eventually leads, by adapting (\ref{mto0}), to the leading correction
\be
J_N^{(2n_1,\cdots ,2n_{N-5})}=2\prod_{l=1}^{N-5}\frac{1}{(2n_l)!}\int\prod_{l=1}^{N-5}\prod_{i_l=1}^{2n_l}\frac{d\alpha_{i_l}^{(l)}}{2\pi}g^{(2n_1,\cdots ,2n_{N-5})}(\alpha_2^{(1)},\cdots,\alpha_{2n_1}^{(1)};\alpha_1^{(2)}\cdots\alpha_{2n_{N-5}}^{(N-5)}) \,.
\ee

\end{document}